\documentclass[preprint,12pt,3p]{elsarticle}%
\usepackage{xurl}
\usepackage{eurosym}
\usepackage{makecell}
\usepackage{siunitx}
\usepackage{amsfonts}
\usepackage{amsmath}
\usepackage{amssymb}
\usepackage{graphicx}
\usepackage{subcaption}
\usepackage{placeins}
\usepackage{subfig}
\usepackage{caption}
\usepackage{soul}
\usepackage[utf8]{inputenc}
\usepackage{xcolor}%
\newcommand{\conv}{\mathbin{\ast}}

\newcommand{\nmR}[1]{\textcolor{black}{#1}}
\newcommand{\nm}[1]{\textcolor{black}{#1}}

\newcommand{\gc}[1]{\textcolor{black}{#1}}
\begin{document}

\begin{frontmatter}

\title{Theory of fracture initiation and propagation in viscoelastic media}

\author[label1,label2,label3]{Giuseppe Carbone\corref{cor1}}
\author[label2]{Cosimo Mandriota}
\author[label2]{Guido Violano}
\author[label2]{Luciano Afferrante}   
\author[label1,label2]{Nicola~Menga}
\cortext[cor1]{Corresponding author: giuseppe.carbone@poliba.it}
\affiliation[label1]{
    organization={School of Physics, Nanjing University of Science and Technology},
    city={Nanjing},
    postcode={210094},
    country={China}
    }
\affiliation[label2]{
    organization={Department of Mechanics, Mathematics and Management, Politecnico di Bari},
    city={Bari},
    country={Italy}
    }
\affiliation[label3]{
    organization={CNR - Institute for Photonics and Nanotechnologies U.O.S. Bari, Physics Department "M. Merlin"},
    city={Bari},
    country={Italy}
    }

\begin{abstract}

Crack initiation and propagation are fundamental problems in materials science, often leading to catastrophic failure. While fracture in elastic solids occurs instantaneously above a critical load, viscoelastic materials may sustain high loads for a finite time before cracks start to propagate. This phenomenon, known as delayed fracture, has been widely observed experimentally but is still only partially understood theoretically. In this study, we present a rigorous framework based on the Lagrange–d’Alembert principle of virtual work (PVW) to predict both the viscoelastic delay time and the subsequent crack evolution under arbitrary loading histories. We derive how the delay time depends on the applied remote load and validate the theory through quantitative comparison with experiments, using directly measured delay times together with DMA-based viscoelastic characterization of the material. Very good agreement is obtained over a broad range of loading and delay times. Our results also show that crack propagation starts at finite speed and that load-dependent steady-state conditions are soon established. Finite element analyses further support the proposed framework and clarify the role of finite-ranged adhesion forces at fixed adhesion energy, showing that shorter interaction ranges yield results in quantitative agreement with theory. We also present, for the first time, a rigorous $J$-integral formulation valid for linear viscoelastic solids under arbitrary, time-varying loading histories. The result restores path independence and yields a generalized Griffith criterion that naturally predicts delayed fracture initiation in non-conservative materials. Remarkably, fracture initiation can be described without specifying the detailed stress distribution within the process zone, as long as it remains small relative to the crack length.

\end{abstract}


\end{frontmatter}

\section{Introduction}

Understanding and predicting fracture initiation and propagation in
viscoelastic media is essential to engineer the mechanical behavior of
polymeric and composites materials, which are of ubiquitous use in
engineering, e.g., in aerospace and automotive technologies
\cite{Awaja2016,Choi2005,Irez2020}. Studying the failure mechanisms of
viscoelastic gels, elastomers and biological tissues is pivotal in the food
industry \cite{TabiloMunizaga2005,BarbosaCnovas1996}, in batteries with solid electrolytes \cite{Ding2021,Zhang2023,Cheng2024}, in drug delivery systems
\cite{Majumder2022,Sree2023}, and many bio-medical fields
\cite{Viano1986,Zioupos1998,Ateshian2022}, where, critical safety conditions
must be often ensured, e.g., in tissue regeneration \cite{Neffe2015}, heart
valves \cite{Simon2003,Jiao2012}, and cardiovascular stents
\cite{Harewood2007,Shanahan2017}. Importantly, for this class of materials, a
critical fracture condition may not be uniquely determined by the magnitude of
a certain remote applied stress, but it depends also on the loading history. This
phenomenon, usually referred to as delayed fracture, has been studied
experimentally, with several contributions published in the past couple of
decades. Delayed failure under a step-change in the remote stress has been
observed in several viscoelastic materials, such as hydrogels
\cite{Karobi2016,Bonn1998,Tang2017,Skrzeszewska2010}, colloidal gels
\cite{Lindstrm2012,Sprakel2011}, protein gels \cite{Brenner2013}, metal alloys
undergoing high temperatures \cite{Kuduzovi2014}, elastomers
\cite{Ju2023,Wang2023,Mishra2018} and various other polymers
\cite{Frassine1996,Frankiewicz1972,Knauss1970,Tabuteau2009}. A common
observation is that the time required for fracture initiation decreases as the
remote stress is increased. Experimental studies have focused on delayed
fracture originating from macroscopic cracks of various geometries
\cite{Frankiewicz1972,Knauss1970,vanderKooij2018,Tang2017}, or from the
nucleation of micro defects induced by tensile \cite{Bonn1998,Karobi2016} or
shear \cite{Brenner2013,Skrzeszewska2010,Sprakel2011} stresses, or by droplets
as well \cite{Grzelka2017}. A delayed failure under apparently subcritical
loads represents an hardly predictable phenomenon, with critical negative
consequences for materials' safety
\cite{Guarino2002,Kun2003,Lindstrm2012,Wang2012}. Nonetheless, in some
applications, it may even be strategically exploited, e.g., in drug delivery
systems, to control the drugs release activated by crack propagation in
polymeric structures \cite{Wang2016}. This phenomenon has been often
attributed to physical-chemical processes at the molecular scale, such as
water migration in hydrogels \cite{Tang2017}, bonds and strands breaking
dynamics \cite{Lindstrm2012,Brenner2013}, thermally activated bond
dissociation \cite{Skrzeszewska2010,Evans1997}, energy barriers for micro
defects nucleation \cite{Bonn1998,Skrzeszewska2010}. However, since a
viscoelastic rheology is a common denominator in aforementioned observations,
continuum mechanics can provide significative insights in understanding the
role of viscoelasticity in affecting the delay in materials' failure, as
suggested by existing studies
\cite{Kaminsky2014,Knauss1970,Schapery1975,Williams1965,Wnuk1970}.

A very rough interpretation can be provided observing that during creep under constant load, the elastic energy stored into the material increases as the material softens. Thus, only after a certain delay time enough energy could be available to propagate the crack. However, Griffith's criterion \cite{Griffith} cannot predict crack propagation in non-conservative materials, and the contribution of irreversible phenomena at the crack tip appears as a necessary step while modeling viscoelastic fracture mechanics \cite{Ciavarella2021,Knauss1970,Persson2005}.

The assumptions of infinitely short range forces
at the crack interface lead to the standard square root stress singularity
\cite{Irwin1957,Persson2005}, which works well for 
elastic materials \cite{Griffith,Irwin1957}, is usually is relaxed when
modeling crack propagation in viscoelastic solids \cite{Ciavarella2021,Hui2022,Schapery1975} by introducing a cohesive zone of finite length (also called 'process
zone') in which the material is assumed to behave plastically
\cite{DasGupta1977,Wnuk1970} or according to generic non-linear constitutive
laws \cite{Schapery1975} as, for instance, 'Lennard-Jones'-like ones
\cite{Greenwood2007} or power laws \cite{Schapery2022}. Nevertheless, even
more complex phenomena might take place at the crack tip, such as the
formation of cavities and chains pull-out, for which exhaustive models
currently are lacking \cite{Persson2005}. Other studies are based on
macroscopic power balance including viscous dissipation
\cite{Persson2005,deGennes1996, Christensen}, although limited to stationary
crack propagation. Following Knauss' review \cite{Knauss2015}, the existing
literature on viscoelastic fracture mechanics, spanning from the 1960s to very
recent contributions
\cite{Schapery2022,Violano2023,Ciavarella2021.2,Ciavarella2022,Persson2021},
has dedicated more attention to steady-state crack growth \cite{Hui2022}
rather than to crack initiation (i.e., delayed fracture) and unsteady
propagation in the presence of generic loading histories.
\gc{Early attempts to extend path-independent integral concepts to transient crack problems
also exist in the literature. In particular, Nilsson \cite{Nilsson1973} introduced a
path-independent integral in the Laplace-transform domain for dynamic crack problems
with inertia, mainly as a computational tool for evaluating transient stress-intensity
factors. However, as already noted in that work, the inverse Laplace-transform of such an integral is not
equal to the energy release rate and therefore does not provide a direct energetic
fracture criterion.}
To the best of the authors' knowledge, most of the existing studies on viscoelastic delayed
fracture in the framework of continuum mechanics date back to the 1960-70s,
with few recent contributions \cite{Shrimali2023.2, Wang2012}. In most cases,
fracture initiation in elastomers from pre-existing cracks was studied adopting cohesive zone models combined with a material failure criterion, such as critical crack
opening displacement \cite{Frankiewicz1972, Wnuk1970} or fracture energy  criteria \cite{Frankiewicz1972,Knauss1970,Schapery1975, Wang2012}. 

\gc{This study aims at providing an exhaustive and versatile new perspective for approaching linear viscoelastic fracture mechanics. A central contribution of the present work is the derivation of a history-dependent energetic fracture criterion for delayed fracture in linear viscoelastic media. In this framework, the stress intensity factor only plays the role of the far-field loading parameter associated with the asymptotic crack-tip fields, whereas the critical condition for fracture initiation follows from an energy balance that explicitly depends on the loading history through the viscoelastic constitutive response. We also derive a rigorous, path-independent $J$-integral formulation in the real time domain, thus providing a
mechanics-based energetic criterion for delayed fracture initiation under arbitrary loading histories.}
In detail, by relying on the Lagrange-d'Alembert virtual work formalism, we initially study delayed fracture in terms of fracture initiation (i.e., crack tip being still), then we study the subsequent unsteady crack propagation (i.e., moving crack tip).
\gc{We remark that the two-dimensional semi-infinite crack setting adopted here should be
understood in the standard fracture-mechanics sense: it is not intended to reproduce all
the geometric details of a real component, but rather to isolate the universal local
asymptotic fields near the crack tip. In this respect, the effect of the global geometry
and boundary conditions is carried by the stress intensity factor, whereas the energetic
structure of the initiation criterion is local in nature.
Finally, our predictions regarding both the delay time required for fracture initiation
and the subsequent crack propagation phase are compared with finite element simulations,
previous models, and experimental results. In particular, to test the fracture-initiation
criterion, we determine the viscoelastic creep function of PTFE through dynamic mechanical analysis (DMA)
 and then postprocess delayed-fracture experiments on cracked specimens under
time-varying tensile loading. The onset of crack propagation is identified by image
postprocessing, which allows the delayed-fracture time to be quantitatively measured and
directly compared with the corresponding theoretical prediction.}

Importantly, the proposed theoretical framework does not require the removal of the stress singularity at the crack tip. Hence, as long as the crack tip process zone is much smaller than other characteristic lengths, our approach avoids detailed descriptions of the local failure mechanisms, which are fully accounted for by the fracture energy. 
The proposed formalism can be regarded as the extension of previous works of some of the authors on adhesive viscoelastic contact mechanics \cite{Carbone2022,Mandriota2024.2,Mandriota2024}.

\section{Fracture initiation \label{sec:initiation}}

\begin{figure}[tbh]
\centering\includegraphics[width=0.95\textwidth]{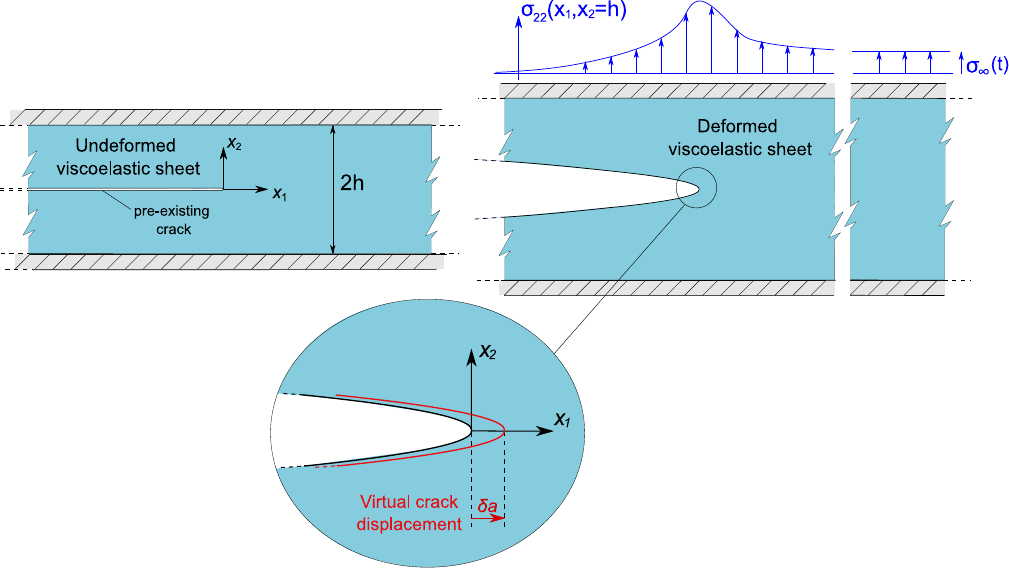}\caption{
\nm{The schematic of the fracture problem: a thin viscoelastic sheet of height $2h$ and infinite horizontal length with a pre-existing semi-infinite crack is put under traction. The top right figure, shows a qualitative time-dependent tensile stress distribution at the clamped sheet boundary, with a far field stress value $\sigma_\infty(t)=\sigma_{22}(x_1 \to+\infty,x_2,t)$. The bottom figure shows the crack tip, with the red line representing the deformed shape associated with a virtual change $\delta a$ of the crack length.}
}%
\label{fig1}%
\end{figure}

In this section, we employ the Principle of Virtual Works (PVW) to derive the
critical condition for fracture initiation in a viscoelastic media. 
\nm{To this end, we consider the system shown in Fig. \ref{fig1}, where a homogeneous and isotropic viscoelastic thin sheet of height $2h$ with a pre-existing semi-infinite straight crack undergoes traction, under plane stress conditions. Notably, the crack tip is standing still, and the controlled quantity is the far field uniaxial stress $\sigma_\infty(t)=\sigma_{22}(x_1 \to+\infty,x_2,t)>0$ at the extreme right-hand side of the sheet.} 

\nm{We define the fixed reference frame $(x_{1},x_{2})$, with the $x_{1}$-axis aligned along the pre-existing crack. For simplicity, we set the origin of $(x_{1},x_{2})$ corresponding to the pre-exisiting crack tip, as shown in Fig. \ref{fig1}; however, we remark that the reference system is never co-moving with the crack tip. Moreover, since the problem is translationally invariant along the $x_1$ axis, should the pre-existing crack tip be located at a generic coordinate $a>0$ (which we call 'crack length' in what follows), all the fields in the sheet would be simply shifted of this quantity. As a consequence, we also have $\partial/\partial a=-\partial/\partial x_1$.}
The stress and strain tensors fields are $\sigma_{ij}\left(  x_{1},x_{2},t\right)$ and $\varepsilon_{ij}\left(x_{1},x_{2},t\right)$, respectively, and the displacement vector field is $\mathbf{u}\left(  x_{1},x_{2},t\right)$, with $i,j=1,2$. 
Moreover, we assume that sufficiently far in the past the sheet was unloaded and undeformed, i.e. $\sigma_{ij}\left(  x_{1},x_{2},-\infty\right)  =0$ and $\varepsilon_{ij}\left(  x_{1},x_{2},-\infty\right)  =0$. Finally, regarding the crack, we assume that the upper and bottom fractured surfaces do not interact with each-other (i.e., no finite range interactions).
In such conditions, the stress field in the media is independent of the detailed relation between the stress and strain for a homogeneous material (i.e., also valid for a linear viscoelastic material) \cite{Carbone2005,carbone2005.2,Christensen,DAmico2013,Graham1973,Graham1970,Mueller, Rivlin1953,Schapery1975,Yeoh2003}\nm{, and is given in general by}
\begin{equation}
\sigma_{ij}\left(  x_{1},x_{2},t\right)  =\sigma_{\infty}\left(  t\right)
\chi_{ij}\left(  x_{1}-a,x_{2}\right)  \label{self-similiarity}%
\end{equation}
where $\chi_{ij}\left(  x_{1}-a,x_{2}\right)  $ are time-independent shape
functions that satisfy the following boundary conditions
\begin{equation}
\chi_{11}\left(  \pm\infty,x_{2}\right)  =0;~\chi_{22}\left(  +\infty
,x_{2}\right)  =1;~\chi_{22}\left(  -\infty,x_{2}\right)  =0;~\chi_{12}\left(
\pm\infty,x_{2}\right)  =0\nonumber
\end{equation}

In viscoelastic materials, \nm{fracture initiation} can be delayed by a (delay) time $t_{\mathrm{d}}$ after applying the tensile stress $\sigma_{\infty}$ to the sheet. 
The PVW allows us to define the \nm{corresponding} critical conditions, i.e. when the virtual work of internal stresses $\delta L_{\mathrm{I}}$ balances the virtual work of external stresses $\delta L_{\mathrm{E}}$ for any virtual variation $\delta a$ of the crack length (i.e., of the Lagrangian parameter $a$) made at fixed time $t$ (i.e., virtually taking a snapshot of the system state at time $t$ and applying the change of $a$ at that state). 
Hence, 
\begin{equation}
\delta L_{\mathrm{I}}(t)=\delta L_{\mathrm{E}}(t) \label{VWP1}%
\end{equation}
where 
\begin{equation}
\delta L_{\mathrm{I}}(t)=\int d^{2}x\sigma_{ij}\left(  x_{1},x_{2},t\right)
\delta\varepsilon_{ij}\left(  x_{1},x_{2},t\right)  \label{interal work}%
\end{equation}
with $\delta\varepsilon_{ij}\left(  x_{1},x_{2},t\right)$ being
the virtual strain field caused by the virtual variation $\delta a$ of the crack length 

Using Eq. (\ref{self-similiarity}) and recalling that for linear viscoelastic solids%
\begin{equation}
\varepsilon_{ij}\left(  x_{1},x_{2},t\right)  =\int_{-\infty}^{t}%
dt_{1}J_{ijhk}\left(  t-t_{1}\right)  \dot{\sigma}_{hk}\left(  x_{1}%
,x_{2},t_{1}\right)  \label{strain visco}%
\end{equation}
with $J_{ijhk}\left(  t\right)  $ being the local fourth order creep
compliance tensor, we get%
\begin{equation}
\varepsilon_{ij}\left(  x_{1},x_{2},t\right)  =\int_{-\infty}^{t}%
dt_{1}J_{ijhk}\left(  t-t_{1}\right)  \chi_{hk}\left(  x_{1}-a,x_{2}\right)
\dot{\sigma}_{\infty}\left(  t_{1}\right)  \label{strain field}%
\end{equation}
and%
\begin{equation}
\delta\varepsilon_{ij}\left(  x_{1},x_{2},t\right)  =-\delta a\frac
{\partial\chi_{hk}(x_{1}-a,x_{2})}{\partial x_{1}}\int_{-\infty}^{t}%
dt_{1}J_{ijhk}\left(  t-t_{1}\right)  \dot{\sigma}_{\infty}\left(
t_{1}\right)  \label{virtual strain 2}%
\end{equation}
\nm{where we used $\partial/\partial a=-\partial/\partial x_1$.} 
Eq.(\ref{interal work}) then gives%
\begin{equation}
\delta L_{\mathrm{I}}(t)=-\delta a\sigma_{\infty}\left(  t\right)
\int_{-\infty}^{t}dt_{1}\dot{\sigma}_{\infty}\left(  t_{1}\right)  \int
d^{2}xJ_{ijhk}\left(  t-t_{1}\right)  \chi_{ij}\left(  x_{1}-a,x_{2}\right)
\frac{\partial\chi_{hk}(x_{1}-a,x_{2})}{\partial x_{1}}
\label{internal work 2}%
\end{equation}

Note that the creep compliance fourth order tensor possesses the major symmetry, meaning that $J_{ijhk}\left(  t\right) = J_{hkij}(t)$. Such property can be obtained from the Green-Kubo fluctuation dissipation theorem and the Onsager time-reversal invariance {(see \ref{sec:app B} for the detailed derivation)}, and for isotropic materials it yields \cite{Christensen}%
\begin{equation}
J_{ijhk}\left(  t\right)  =\alpha_{1}\left(  t\right)  \delta_{ij}\delta_{hk}+\alpha_{2}\left(t\right)  \left(  \delta_{ih}\delta_{jk}+\delta_{ik}\delta_{jh}\right)
\label{Christensen}%
\end{equation}
By using the major symmetry property we write
\begin{equation}
\begin{split}
&  J_{ijhk}\left(  t-t_{1}\right)  \chi_{ij}\left(  x_{1}-a,x_{2}\right)
\frac{\partial\chi_{hk}(x_{1}-a,x_{2})}{\partial x_{1}}= \\
&\frac{1}{2}%
\frac{\partial}{\partial x_{1}}\left[  J_{ijhk}\left(  t-t_{1}\right)
\chi_{ij}\left(  x_{1}-a,x_{2}\right)  \chi_{hk}(x_{1}-a,x_{2})\right]
\end{split}
\label{passaggio derivata}
\end{equation}
Finally, replacing Eq. (\ref{passaggio derivata}) in Eq. (\ref{internal work 2}), we get%
\begin{equation}
\delta L_{\mathrm{I}}(t)=-\delta ah\sigma_{\infty}\left(  t\right)
\int_{-\infty}^{t}dt_{1}J\left(  t-t_{1}\right)  \dot{\sigma}_{\infty}\left(
t_{1}\right)  =-\delta ah\sigma_{\infty}\left(  t\right)  \varepsilon_{\infty
}\left(  t\right)  \label{internal work final}%
\end{equation}
where $J\left(  t\right)  =J_{2222}\left(  t\right)  $ is the uniaxial creep
function of the viscoelastic material, and 
\begin{equation}
\varepsilon_{22}\left(
+\infty,x_{2},t\right)  =\varepsilon_{\infty}\left(  t\right)  =\int_{-\infty
}^{t}dt_{1}J\left(  t-t_{1}\right)  \dot{\sigma}_{\infty}\left(  t_{1}\right)
\label{eps_infty}    
\end{equation}
is the remote strain.

The above equations allows us to define the energy release rate $G\left(
t\right)  $ as%
\begin{equation}
G\left(  t\right)  =-\frac{\delta L_{\mathrm{I}}(t)}{\delta a}=h\sigma
_{\infty}\left(  t\right)  \varepsilon_{\infty}\left(  t\right)
\label{general energy release rate}%
\end{equation}
Focusing on \nm{the virtual work of external stresses} $\delta L_{\mathrm{E}}(t)$ in Eq. (\ref{VWP1}), following the PVW, the (admissible) virtual displacements field $\delta u_{i}\left(  x_{1},x_{2},t\right)  $, $i=1,2$ is required to satisfy the compatibility conditions
\begin{equation}
\delta\varepsilon_{ij}(x_{1},x_{2},t)=\frac{1}{2}\left(  \frac{\partial\delta
u_{i}}{\partial x_{j}}+\frac{\partial\delta u_{j}}{\partial x_{i}}\right)
\label{compatibility}%
\end{equation}
However, $\delta u_{i}\left(  x_{1},x_{2},t\right)$ can be easily derived by shifting of the quantity $\delta a$ the original displacement fields $u_{i}\left(  x_{1},x_{2},t\right) $ associated with the strain fields $\varepsilon_{ij}(x_{1},x_{2},t)$; therefore,
\begin{equation}
\delta u_{i}\left(  x_{1},x_{2},t\right)  =-\delta a\frac{\partial
u_{i}\left(  x_{1},x_{2},t\right)  }{\partial x_{1}}
\label{virtual displacement}%
\end{equation}
Noting that that $\delta u_{2}\left(  \pm h,x_{2}\right) =h\delta\varepsilon_{\infty}=0$ (no work is associated with the remote stress $\sigma_{\infty}$), the virtual work of external stress is only related to the crack opening and, in turn, to the fracture energy $\Delta\gamma$, i.e.%
\begin{equation}
\delta L_{\mathrm{E}}(t)=-2\Delta\gamma\delta a \label{external work}%
\end{equation}
Finally, combining Eqs. (\ref{VWP1},\ref{general energy release rate},\ref{external work}), we get the critical condition to trigger the crack \nm{initiation}, i.e.%
\begin{equation}
G\left(  t\right)  =h\sigma_{\infty}\left(  t\right)  \varepsilon_{\infty
}\left(  t\right)  =2\Delta\gamma
\label{general critical condition for anysotropic material}%
\end{equation}
Since for homogeneous and isotropic materials the stress intensity factor $K_{\mathrm{I}%
}\left(  t\right)  =\sigma_{\infty}\left(  t\right)  \sqrt{h}$, Eq. (\ref{general energy release rate}) can be rewritten using Eq. (\ref{eps_infty}) as%
\begin{equation}
G\left(  t\right)  =h\sigma_{\infty}\left(  t\right)  \varepsilon_{\infty
}\left(  t\right)  =K_{\mathrm{I}}\left(  t\right)  \int_{-\infty}^{t}%
dt_{1}J\left(  t-t_{1}\right)  \dot{K}_{\mathrm{I}}\left(  t_{1}\right)
\label{energy release rate}%
\end{equation}
and Eq. (\ref{general critical condition for anysotropic material}) yields%
\begin{equation}
G(t) = K_{\mathrm{I}}\left(  t\right)  \int_{-\infty}^{t}dt_{1}J\left(
t-t_{1}\right)  \dot{K}_{\mathrm{I}}\left(  t_{1}\right)  =2\Delta
\gamma\label{critical condition for crack propagaion}%
\end{equation}
which represents a generalization of the Griffith criterion to isotropic homogeneous viscoelastic media.
It suggests that, for a viscoelastic material, crack propagation starts after a certain delay time $t_{\mathrm{d}}$, i.e. when $G\left(  t_{\mathrm{d}}\right)  =2\Delta\gamma$, which depends on the loading history and can be calculated solving Eq. (\ref{critical condition for crack propagaion}).

\gc{It is important to stress that Eq. (\ref{critical condition for crack propagaion}) is not an equation for calculating the stress intensity factor. Rather, $K_I(t)$ enters the formulation as the far-field loading
parameter associated with the asymptotic crack-tip fields, while Eq. (\ref{critical condition for crack propagaion}) determines whether, and at what time, the history-dependent energy balance required for fracture
initiation is satisfied. 
\nmR{In this sense, the convolution product in Eq. (\ref{critical condition for crack propagaion}) shows that this energy balance depends on the entire loading history, and that the critical condition for crack propagation is energetic in nature and, for viscoelastic materials, cannot be simply associated to a critical value $K_I = K_{Ic}$  of stress intensity factor.}
In \ref{sec:app J-int} we also present an extended $J$-integral formulation for viscoelastic
materials that, when evaluated along any path $\Gamma$ surrounding the crack tip [see
Eqs. (\ref{J integral}) and (\ref{J integral2})], leads to the same fracture criterion as Eq. (\ref{critical condition for crack propagaion}). Importantly, the present $J$-integral is formulated directly in the real time domain and preserves path independence even in the case of non-conservative viscoelastic materials. This makes it
particularly useful for both experimental measurements (e.g., digital image correlation)
and numerical analyses (e.g., FEM).} The $J$-integral can be directly evaluated along arbitrary contours surrounding the crack tip, regardless of the detailed stress distribution within the process zone. Furthermore, Eq. (\ref{critical condition for crack propagaion}) can also be derived by enforcing the local equilibrium conditions at the crack tip, as reported in Ref. \cite{Mandriota2024}, following the procedure shown in \ref{sec:app A}.
\nm{The energy release rate} $G(t)$, as defined in Eq.
(\ref{general critical condition for anysotropic material}) or Eq.
(\ref{energy release rate}), is the work of internal stresses per unit virtual
displacement of the crack tip, which includes the contribution of both the change of the stored
elastic energy and the virtual work of viscous stresses far from the crack
tip. In this respect, it differs from the common elastic energy release rate
$G_{\mathcal{E}}=-\partial\mathcal{E}/\partial a$ (usually defined for elastic
materials) which only represents the release rate of elastic energy per unit
virtual increase of crack length. In the case of isotropic elastic materials
with Young's modulus $E$, being $J\left(  t\right)  =\mathcal{H}(t)/E$ where
$\mathcal{H}(t)$ is the Heaviside unit step function, and equivalently
$\varepsilon_{\infty}\left(  t\right)  =\sigma_{\infty}\left(  t\right)  /E$
we recover the classical expression%
\begin{equation}
G\left(  t\right)  =\frac{K_{\mathrm{I}}^{2}\left(  t\right)  }{E}%
=\frac{h\sigma_{\infty}^{2}\left(  t\right)  }{E}=G_{\mathcal{E}}\left(
t\right)  \label{elastic energy release rate}%
\end{equation}

In what follows, we consider the effect of different loading histories on the \nm{delay time $t_\mathrm{d}$}. To simplify the treatment, we assume that the viscoelastic material is characterized by a single relaxation time $\tau$, unless otherwise specified, with the creep function $J(t)$ given by
\begin{equation}
J(t)=\mathcal{H}\left(  t\right)  \left\{  \frac{1}{E_{\infty}}+\left(
\frac{1}{E_{0}}-\frac{1}{E_{\infty}}\right)  \left(  1-e^{-t/\tau}\right)
\right\}  \label{creep function}%
\end{equation}
where $E_{\infty}$ and $E_{0}$ are, respectively, the high frequency and the
low frequency viscoelastic moduli, and $\tau$ is the characteristic
retardation time of the system. However, we remark that the presented methodology is general and can be applied to viscoelastic materials with relaxation times spanning over several decades.

\subsection{Step load\label{step load}}

Let us assume that the remote applied stress undergoes a step change from zero
to a finite value as $\sigma_{\infty}$ is applied remotely at time $t=0$, i.e.
$\sigma_{\infty}\left(  t\right)  =\sigma_{\infty}\mathcal{H}\left(  t\right)
$ and the stress intensity factor is $K_{\mathrm{I}}\mathcal{H}\left(
t\right)  $ with $K_{\mathrm{I}}=\sigma_{\infty}\sqrt{h}$ and $\mathcal{H}%
\left(  t\right)  $ being the Heaviside unit step function. Using the generalized
Griffith condition, as reported in Eq.
(\ref{critical condition for crack propagaion}) the delay time $t_{\mathrm{d}}$, at the onset of crack propagation can be determined by solving the
equation
\begin{equation}
G(t_{\mathrm{d}})=J\left(  t_{\mathrm{d}}\right)  K_{\mathrm{I}%
}^{2}=2\Delta\gamma\label{General Grifft Visco}%
\end{equation}

\begin{figure}[ptb]
\centering\includegraphics[width=0.99\textwidth]{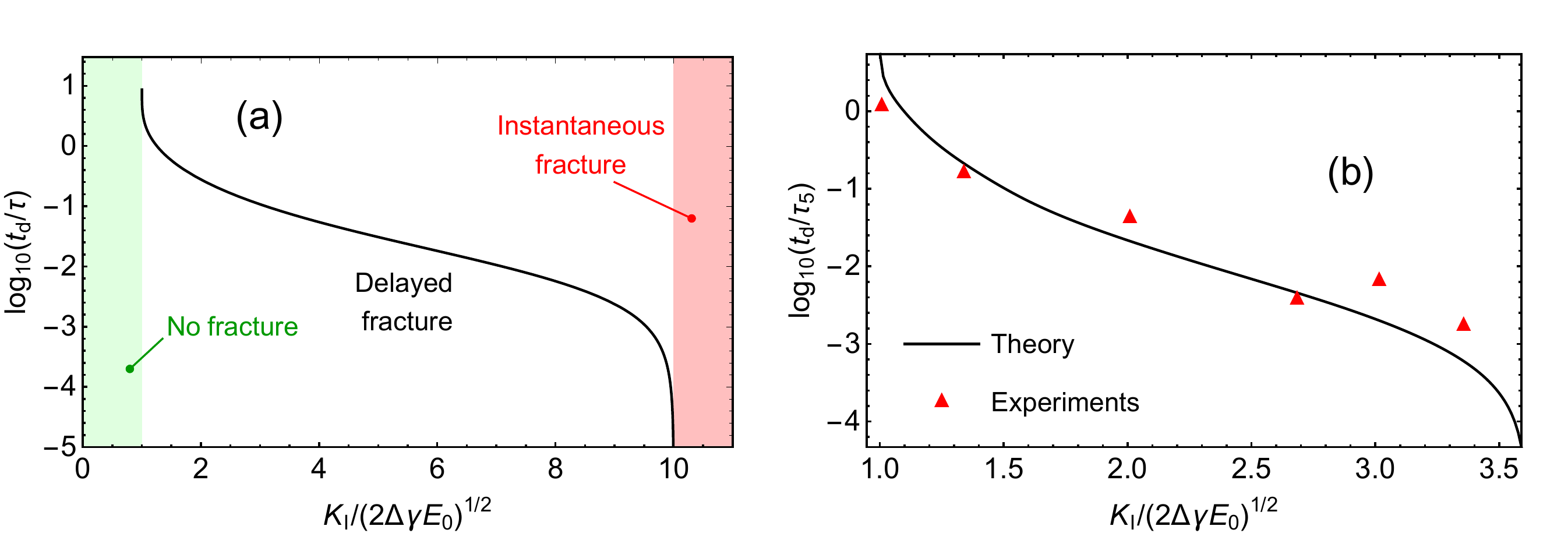}\caption{{The delay time in response to a step load. (a): the dimensionless delay time $t_{\mathrm{d}}/\tau$ as function of the dimensionless stress intensity factor magnitude $K_{\mathrm{I}}/(2E_{0}\Delta\gamma)^{1/2}$. Results are provided for $E_{\infty}/E_{0}=100$. (b) Dimensionless comparison between theoretical predictions and experimental data from Ref. \cite{Karobi2016} - see the text for details.}}%
\label{figDelayTimestep}%
\end{figure}

\begin{figure}[ptb]
\centering\includegraphics[width=0.5\textwidth]{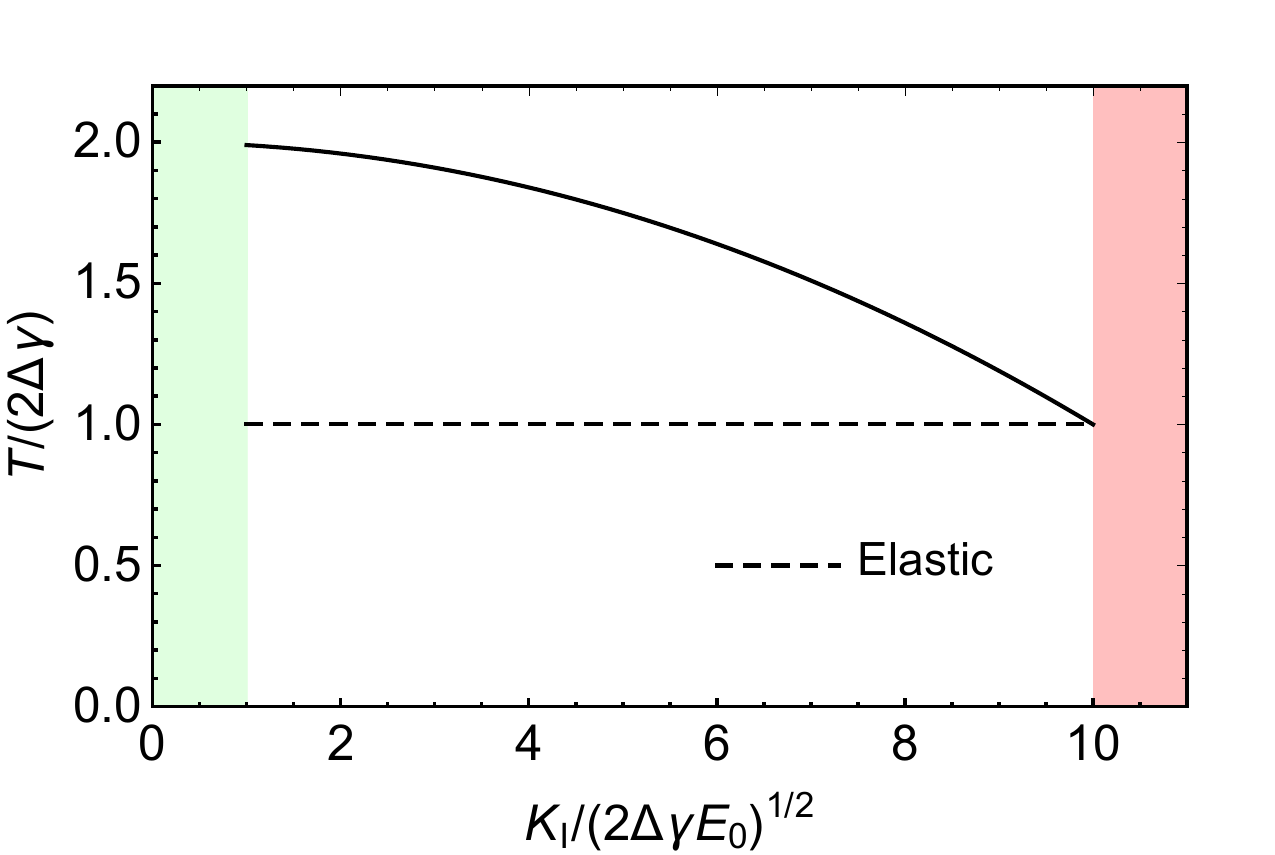}\caption{{The dimensionless toughness $T/\left(2\Delta\gamma\right)$ as function of the dimensionless stress intensity factor magnitude $K_{\mathrm{I}}/\left(2\Delta\gamma E_{0}\right)^{1/2}$, resulting from a step load. Results are provided for $E_{\infty}/E_{0}=100$. The dashed line is the corresponding elastic case according to the Griffith's fracture criterion for purely elastic materials.}}%
\label{figToughness}%
\end{figure}
\FloatBarrier

The quantity $t_{\mathrm{d}}$ clearly depends on the specific form of the
viscoelastic creep function. However, since $J(t)$ is a monotonic increasing
function of time, Eq. (\ref{General Grifft Visco}) admits a real solution
provided that the stress intensity factor $K_{\mathrm{I}}$
\begin{equation}
\left(  2E_{0}\Delta\gamma\right)  ^{1/2}\leq K_{\mathrm{I}}\leq\left(
2E_{\infty}\Delta\gamma\right)  ^{1/2}\label{inequality}%
\end{equation}
Indeed, in response to a step load, the viscoelastic material
behaves elastically with elastic modulus $E_{\infty}$ for $t=0^{+}$; therefore, using
Eq. (\ref{critical condition for crack propagaion}), instantaneous fracture
occurs if $K_{\mathrm{I}}>\left(  2E_{\infty}\Delta\gamma\right)  ^{1/2}$. On
the other hand, when $K_{\mathrm{I}}<\left(  2E_{0}\Delta\gamma\right)
^{1/2}$ crack propagation cannot take place even in the very long-term limit, when the material is fully relaxed
and responds as a soft elastic material with elastic modulus
$E_{0}$. For intermediate values of $K_{\mathrm{I}}$ falling in the range
defined by Eq. (\ref{inequality}), crack propagation starts after a
certain delay time $t_{\mathrm{d}}$, which exponentially decreases with
the stress intensity factor $K_{\mathrm{I}}$ (i.e., with the applied stress $\sigma_\infty$), as shown in Fig. \ref{figDelayTimestep}(a). This is in
perfect agreement with the experimental observations reported in Ref.
\cite{Bonn1998,Skrzeszewska2010,Karobi2016,Brenner2013,Lindstrm2012,Mishra2018}
as also confirmed in Fig. \ref{figDelayTimestep}(b), where theoretical predictions are
compared with experimental data presented in Ref. \cite{Karobi2016}.

The latter experiments \cite{Karobi2016} consist of several tensile creep tests on dumbbell-shaped specimens, where fracture occurs because of the nucleation and growth of 
subcritical microcracks. Regardless of the specific geometry of the microscopic crack, the stress intensity factor scales proportionally to the applied stress. Therefore, for each test we get $K_{\mathrm{I}}=\langle K_{\mathrm{I}}\rangle\sigma_{\infty}/\langle\sigma_{\infty}\rangle$ where "$\langle\rangle$" denotes the average stress intensity factor. 
Thus, according to Eq. (\ref{critical condition for crack propagaion}) the only parameter needed to fit the experimental data is the dimensionless average stress intensity factor $\langle K_{\mathrm{I}}\rangle /(2\Delta\gamma E_{0})^{1/2}$. 
Focusing on the experimental fracture data
\footnote{Data in Ref. \cite{Karobi2016} have been fitted using $J(t)=\mathcal{H} \left(  t\right)  \left\{E_{\infty}^{-1}+\sum_{k=1}^{5}E^{-1}\left[1-\exp\left(  -t/\tau_{k}\right)  \right]  \right\}  $ for $\tau_k$ and $E_k$ according to the creep test at $\sigma_{\infty}=0.2~\mathrm{MPa}$ on the gel denoted c-PA1. We found $\tau_{k}=\{\,26;\;156;\;939;\;5637;\;33825\,\}\,\mathrm{s}$, $E_{k}=\{\,6.3;\;1.1;\;0.22;\;0.15;\;0.08\,\}\,\mathrm{MPa}$ and $E_{\infty}=0.49~\mathrm{MPa}$. Consequently, we have $E_{0}=0.037~\mathrm{MPa}$.} 
of gel c-PA1, we get $\left\langle \mathit{K_{\mathrm{I}}}\right\rangle /(2\Delta\gamma E_{0})^{1/2}\approx2.24$. Then, using this value, we can determine the delay time $t_\mathrm{d}$ across the entire range of the stress intensity factor $K_{\mathrm{I}}$, obtaining a very good agreement with the experimental data.

Viscoelasticity also affects the system toughness $T$, i.e. the work per unit
surface done by the remote stress $\sigma_{\infty}$ up to the onset of crack
propagation. Hence, the dimensionless toughness $T/\left(  2\Delta\gamma\right)  $
is%
\begin{equation}
\frac{T}{2\Delta\gamma}=\frac{h}{\Delta\gamma}\int_{-\infty}^{t_{\mathrm{d}}%
}dt\sigma_{\infty}\left(  t\right)  \dot{\varepsilon}_{\infty}\left(
t\right)  =2-\frac{K_{\mathrm{I}}^{2}}{2E_{\infty}\Delta\gamma}=2-\frac
{1}{J\left(  t_{\mathrm{d}}\right)  E_{\infty}}\geq1 \label{toughness}%
\end{equation}
where we have used that $K_{\mathrm{I}}^{2}J\left(  t_{\mathrm{d}}\right)
=2\Delta\gamma$ as required by the equilibrium condition Eq.
(\ref{General Grifft Visco}). Note that, because of the hysteretic behavior of
the viscoelastic material, $T/\left(  2\Delta\gamma\right)  \geq1$. More
specifically only for $K_{\mathrm{I}}=\left(  2E_{\infty}\Delta\gamma\right)
^{1/2}$, i.e. when $t_{\mathrm{d}}\rightarrow0^{+}$ we have as expected
$T/\left(  2\Delta\gamma\right)  =1$, whereas for $K_{\mathrm{I}}=\left(
2E_{0}\Delta\gamma\right)  ^{1/2}$, i.e. when $t_{\mathrm{d}}\rightarrow
\infty$ we get $T/\left(  2\Delta\gamma\right)  =2-E_{0}/E_{\infty}\approx2$,
where we have used that $E_{\infty}\gg E_{0}$. Moreover, Eq. (\ref{toughness})
shows that $T$ must decrease parabolically with the remote stress
$\sigma_{\infty}$ (or equivalently the stress intensity factor $K_{\mathrm{I}%
}$) as clearly shown in Fig. \ref{figToughness}. Such dependence on the applied
stress paves the way to the development of possible stress based techniques
able to tune the effective adhesion strength, stickiness and toughness of
viscoelastic materials
 \cite{Mandriota2024,Mandriota2024.2,Charmet1996,Afferrante2022,Lorenz2013}.

Finally, we remark that when remote strain $\varepsilon_{\infty
}\left(  t\right)  $ is controlled instead of the remote stress, the critical
condition for fracture initiation given by Eq. (\ref{general critical condition for anysotropic material}), i.e. $h\varepsilon_{\infty}\left(  t\right)
\sigma_{\infty}\left(  t\right)  \geq2\Delta\gamma$, reads%
\begin{equation}
h\varepsilon_{\infty}\left(  t\right)  \int dt_{1}R\left(  t-t_{1}\right)
\dot{\varepsilon}_{\infty}\left(  t_{1}\right)  \geq2\Delta\gamma
\label{controlled strain}%
\end{equation}
where $R\left(  t\right)  =R_{2222}\left(  t\right)  $ is the uniaxial
relaxation function of the material. Assuming a step change of the remote
strain $\varepsilon_{\infty}(t)=\varepsilon_{\infty}\mathcal{H}(t)$, the above
condition yields $h\varepsilon_{\infty}^{2}R\left(  t\right)  \geq
2\Delta\gamma$. Importantly, since $R\left(  t\right)  $ is a monotonically decreasing
function of time, delayed fracture cannot occur. Instead, crack propagation can only start immediately for
$h\varepsilon_{\infty}^{2}R\left(  0^{+}\right)  =h\varepsilon_{\infty}%
^{2}E_{\infty}\geq2\Delta\gamma$, i.e.  $\varepsilon_{\infty}\geq
\sqrt{2\Delta\gamma/\left(  hE_{\infty}\right)  }$, and it possibly stops at later
time if $\varepsilon_{\infty}<\sqrt{2\Delta\gamma/\left(  hE_{0}\right)  }$.

\subsection{Saturated ramp load\label{ramp load}}

In this section, we assume that the stress intensity factor increases linearly with time, from zero to a maximum value $K_{\mathrm{I}}^{0}=\kappa t_{0}$, and then stays constant. More in detail, the saturated ramp starts at time $t=0$, while $t_{0}$ is the time at which the maximum value is reached. 
In this case, the time derivative of the stress intensity factor takes the form $\dot{K}_{\mathrm{I}}\left(  t\right)  =\kappa\mathcal{H}(t)\mathcal{H}(t_{0}-t)$, with $\kappa>0$. Using Eq. (\ref{critical condition for crack propagaion}), the critical condition for calculating the delay time $t_{\mathrm{d}}$ is
\begin{equation}
\kappa K_{\mathrm{I}}\left(  t\right)  \int_{0}^{\min\left(  t,t_{0}\right)
}dt_{1}J\left(  t-t_{1}\right)  =2\Delta\gamma
\label{equilibrium condition ramp}%
\end{equation}

\begin{figure}[ptb]
\centering\includegraphics[width=0.5\textwidth]{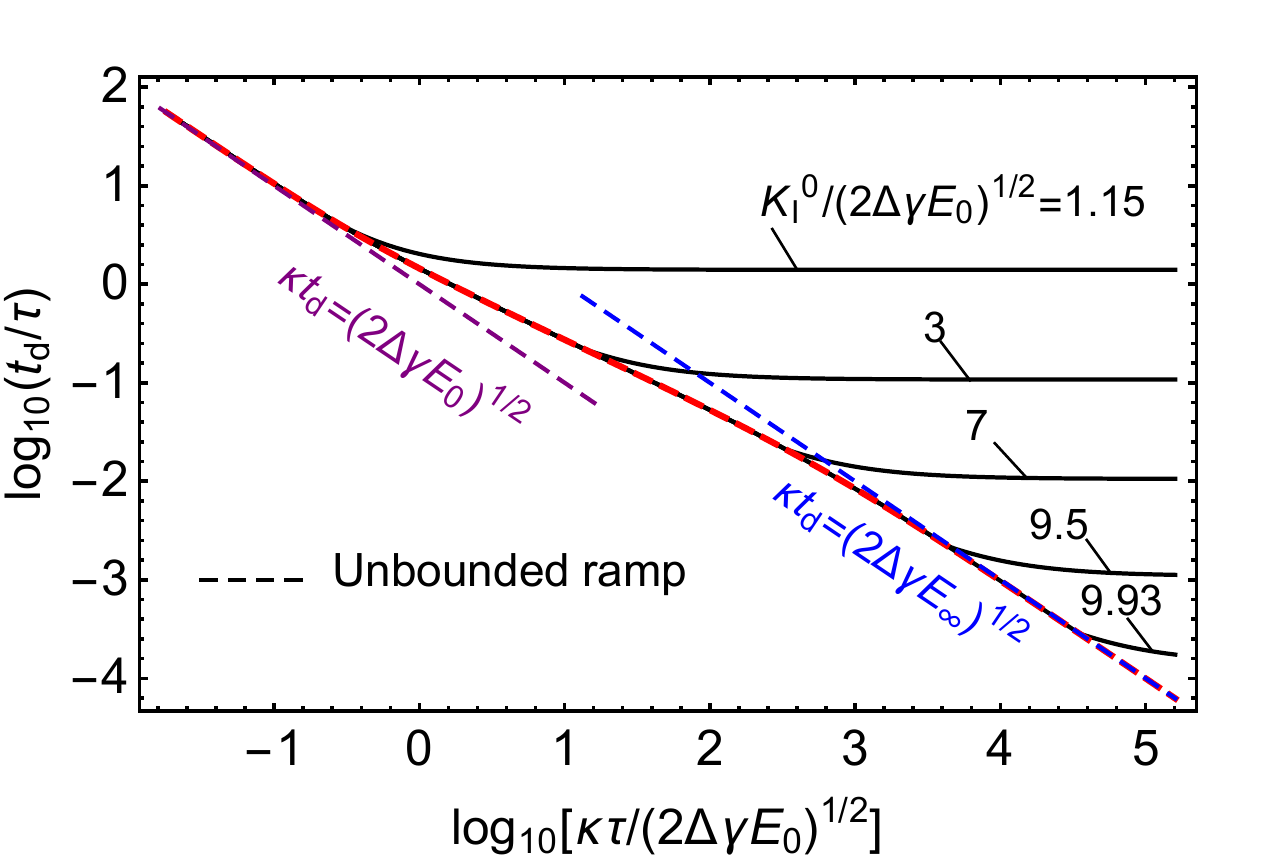}\caption{{Response to stress intensity factor ramping in time as $K_{\mathrm{I}}(t)=\kappa t\mathcal{H}(t)$, up to a final plateau $K_{\mathrm{I}}^{0}$: the dimensionless delay time $t_{\mathrm{d}}/\tau$ as function of the dimensionless ramp slope $\kappa\tau/\left(2\Delta\gamma E_{0}\right)^{1/2}$ for different values of the dimensionless stress intensity factor plateau value $K_{\mathrm{I}}^{0}/\left(2\Delta\gamma E_{0}\right)^{1/2}$. The red dashed line is the delay time in response to the unbounded ramp [$K_{\mathrm{I}}(t)=\kappa t\mathcal{H}(t)$] with asymptotes reported as purple and blue lines. }}%
\label{figDelayTimeRamp}%
\end{figure}

Fig. \ref{figDelayTimeRamp} reports the dimensionless delay time $t_{\mathrm{d}}/\tau$ for
different values of the dimensionless stress intensity factor $K_{\mathrm{I}%
}^{0}/\left(  2\Delta\gamma E_{0}\right)  ^{1/2}$ as a function of $\kappa
\tau/\left(  2\Delta\gamma E_{0}\right)  ^{1/2}$. 

As expected, increasing $K_{\mathrm{I}}^{0}$ at given $\kappa$ leads to shorter values of the delay time. Furthermore, for any given $K_{\mathrm{I}}^{0}\leq\left( 2E_{\infty}\Delta\gamma\right)^{1/2}$, at sufficiently large values of $\kappa$ (i.e. for $t_{0}\ll\tau$) the response to the step load is recovered and the delay time becomes independent of $\kappa$ (see horizontal asymptotes in the right part of Fig. \ref{figDelayTimeRamp}) eventually leading to istantaneously elastic crack propagation for $K_{\mathrm{I}}^{0}=\left( 2E_{\infty}\Delta\gamma\right)^{1/2}$

On the other hand, for sufficiently long or unbounded ramps (i.e., for $t_{0}\gg\tau$) the delay time depends only on the value of $\kappa$, and two asymptotic behaviors are observed (see dashed lines in Fig. \ref{figDelayTimeRamp}). 
The first refers to $\kappa\ll\left(2\Delta\gamma E_{0}\right)^{1/2}/\tau$, when a soft elastic response occurs with modulus $E_{0}$; the second is for $\kappa\gg\left(2\Delta\gamma E_{0}\right)^{1/2}/\tau$ and stiff elastic response with modulus $E_{\infty}$. Using Eq. (\ref{equilibrium condition ramp}) in the elastic limit, crack propagation starts, respectively, for 
\begin{equation}
\kappa t_{\mathrm{d}}=K_I=\sqrt{2\Delta\gamma E_{0}}
\label{asymptote 1}%
\end{equation}
and
\begin{equation}
\kappa t_{\mathrm{d}}=K_I=\sqrt{2\Delta\gamma E_{\infty}}
\label{asymptote 2}%
\end{equation}
In a log-log diagram (see Fig. \ref{figDelayTimeRamp}), both asymptotes are represented by slant lines with slope $-1$ and different intercepts. 

At intermediate values of $\kappa$, the delay time is mainly affected by the viscoelastic hysteresis of the material and the $t_{\mathrm{d}}\ $vs. $\kappa$ curves obtained for different values of $K_{\mathrm{I}}^0$ necessarily collapse on a single master curve as long as $t_{\mathrm{d}}<t_0$, i.e., the solution for the unbounded ramp with $t_0 \rightarrow \infty$ (see red dashed line in Fig. \ref{figDelayTimeRamp}).

\subsection{Power law varying load}
\begin{figure}[ptb]
\centering\includegraphics[width=0.55\textwidth]{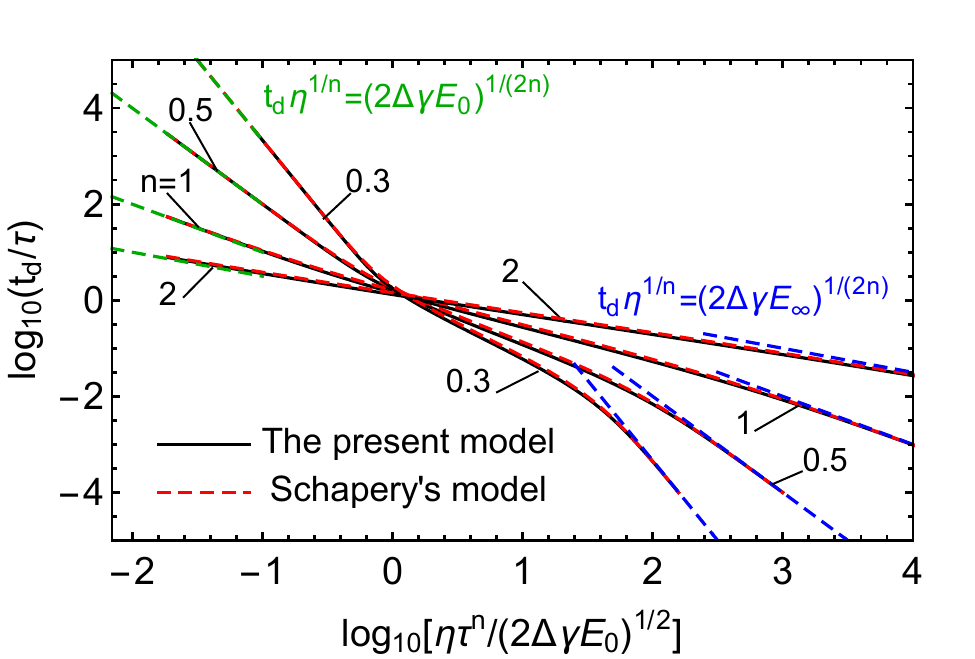}\caption{{
Response to power law varying stress intensity factor $K_{\mathrm{I}}(t)=\eta t^{n}\mathcal{H}(t)$. The dimensionless delay time $t_d/\tau$ as a
function of the dimensionless power law loading pre-factor $\eta\tau^{n}/(2\Delta\gamma E_{0})^{1/2}$. Green and blue dashed lines represent the asymptotic limits and the red dashed line is the corresponding solution predicted by Schapery's model (see Ref. \cite{Schapery1975.2}). Different curves refer to different values
of the power law exponent $n$, and results are given for $E_{\infty}/E_{0}=100$. }}%
\label{figSchapery}%
\end{figure}

In this section, we investigate the delay time corresponding to a stress intensity factor varying as a time power law, i.e. $K_{\mathrm{I}}(t)=\eta t^{n}\mathcal{H}(t)$. Again, according to Eq.
(\ref{critical condition for crack propagaion}), we can identify the two asymptotic limits corresponding to elastic material behaviors elastically, depending on the power law exponent $n$. For very slowly increasing loading, i.e., for $\eta\ll(2\Delta\gamma E_{0})^{1/2}/\tau^{n}$, we
get
\begin{equation}
t_{\mathrm{d}}=\frac{(2\Delta\gamma E_{0})^{1/(2n)}}{\eta^{1/n}}%
\end{equation}
while for fast increases, i.e., for $\eta\gg\tau^{n}/(2\Delta\gamma E_{0})$, we obtain
\begin{equation}
t_{\mathrm{d}}=\frac{(2\Delta\gamma E_{\infty})^{1/(2n)}}{\eta^{1/n}}%
\end{equation}
This is clearly shown in Fig. \ref{figSchapery}, where we also observe a gradual transition (i.e., the intermediate viscous regime) between the two asymptotic limits.

In Fig. \ref{figSchapery}, we also report (dashed red lines) the results from the Schapery approximate formula \cite{Schapery1975,Schapery1975.2}
\begin{equation}
\int_{-\infty}^{t}dt_{1}J\left(  t-t_{1}\right)  \frac{dK_{\mathrm{I}}%
^{2}\left(  t_{1}\right)  }{dt_{1}}=2\int_{-\infty}^{t}dt_{1}J\left(
t-t_{1}\right)  K_{\mathrm{I}}\left(  t_{1}\right)  \dot{K}_{\mathrm{I}%
}\left(  t_{1}\right)  =2\Delta\gamma\label{Shapery}%
\end{equation}
which was obtained under the following two key assumptions: (i) the creep compliance is approximately a power law, (ii) a Dugdale-like cohesive model describes the stress distribution within the crack tip process zone. 
In this regard, we observe that the cohesive stress in Schapery's model is gap-independent, which implies that the size of the process zone is load-dependent (specifically, quadratic - see Eq. (60) in \cite{Schapery1975.2}); as a consequence, the crack tip must move just becasue of a load increase, yet before crack propagation occurs.
We also observe that while our model and Shapery's one provide a formally identical solution in the case of step load, for other cases (e.g., ramp and power law loads) Shapery's formula yields different outcomes, although the numerical implementation of Shapery's approach seems to lead to quantitative small differences (see Fig. \ref{figSchapery}). 
This can be explained by noticing that, at low and high rates $\dot{K}_{\mathrm{I}}\left(  t\right)  $, Eqs. (\ref{Shapery}) and (\ref{critical condition for crack propagaion}) must necessarily provide the same asymptotic elastic results. 
Therefore, in the intermediate range of $\dot{K}_{\mathrm{I}}\left(  t\right)  $ one expects the two approaches to be sufficiently close.
Indeed, assuming a linear increase of $K_{\mathrm{I}}\left(  t\right)  $, one can approximately assume $K_{\mathrm{I}}\left(  t_{1}\right)  \approx K_{\mathrm{I}}\left(  t\right)  /2$ in Eqs. (\ref{Shapery}) thus leading to our Eq. (\ref{critical condition for crack propagaion}). 

Other research groups \cite{Frankiewicz1972, Knauss1970,Wnuk1970} have also investigated delayed fracture in viscoelastic materials using energy calculations. Though different, their formulations are based on the assumption of a non-vanishing size of the crack tip process zone; nonetheless, they all provide results similar to Schapery and the present theory. This allows us to conclude that an accurate description of the stress distribution within the process zone is not a key factor in determining the critical conditions at the onset of crack propagation. This is, indeed, clearly caught by the present theory, which only prescribes that the process zone is sufficiently small compared to other length scales of the problem (note this assumption is always satisfied for the case of a semi-infinite crack in an infinite media).

As a last note, we observe that Eq.(\ref{critical condition for crack propagaion}), and those derived by Schapery and others \cite{Schapery1975,Schapery1975.2,Knauss1970,Frankiewicz1972,Wnuk1970}, are fundamentally different from the theory proposed by Shrimali and Lopez-Pamies \cite{Shrimali2023,Shrimali2023.2}, who suggested that the critical condition at the onset of crack propagation only requires that the change of elastic energy stored in the eventually (fully) relaxed material exactly balances the fracture energy. We observe that their criterion does not follow from the principle of virtual work (i.e. from energy balance - see also \ref{sec:app C}) and may hold true only when a critical threshold strain value is adopted as fracture criterion at the crack tip.

\FloatBarrier

\section{Crack propagation \label{sec:propagation}}

So far we have focused on the critical condition for \nm{fracture initiation}, which allowed us to calculate the delay time $t_{\mathrm{d}}$ depending on the load history. 
In this section, we study \nm{the crack propagation (i.e., moving crack tip) for} $t>t_{\mathrm{d}}$, aiming to determine the time history $a\left(  t\right)$ of the crack length. 
Therefore, we follow the approach already presented by some of the authors in Ref. \cite{Mandriota2024}, i.e. apply the PVW locally at the crack tip, leading to the equilibrium condition
\begin{equation}
\frac{1}{2\delta a}\int_{a}^{a+\delta a}dx_{1}\sigma_{22}%
(x_{1},0,t)u_{2}(x_{1}-\delta a,0,t)=\Delta\gamma
\label{crack tip work balance}%
\end{equation}

\gc{In this respect, the present local formulation goes beyond a pure initiation criterion,
since it provides a framework to address unsteady crack propagation in a viscoelastic
non-conservative medium.}
When crack propagates, for $t>t_{\mathrm{d}}$, Eq. (\ref{crack tip work balance}) requires to specify the stress and displacement distributions at the crack tip. Since the asymptotic stress field close to the crack \cite{Carbone2005,carbone2005.2, Christensen, Schapery1975} is%
\begin{equation}
\sigma_{22}(x_{1},0,t)=\frac{K_{\mathrm{I}}(t)}{\sqrt{2\pi\left[
x_{1}-a(t)\right]  }}\mathcal{H}[x_{1}-a(t)]\label{squarer}%
\end{equation}
using the extended correspondence principle \cite{Christensen,Schapery1984,Schapery1975} allows one to calculate also the local displacement field as%
\begin{align}
u_{2}(x_{1},0,t) &  =J(0^{+})K_{\mathrm{I}}(t)\mathcal{H}\left[
a(t)-x_{1}\right]  \sqrt{\frac{8}{\pi}\left[  a(t)-x_{1}\right]
}
\nonumber
\\
&  +\int_{-\infty}^{t}dt_{1}\dot{J}(t-t_{1})K_{\mathrm{I}}(t_{1}%
)\mathcal{H}\left[  a(t_{1})-x_{1}\right]  \sqrt{\frac{8}{\pi}\left[
a(t_{1})-x_{1}\right]  }
\label{disp2}
\end{align}

Finally, combining Eqs. (\ref{crack tip work balance}-\ref{disp2}), we get
\begin{align}
&  \int_{a}^{a+\delta a}dx_{1}\frac{K_{\mathrm{I}}(t)}{\sqrt{2\pi\left[
x_{1}-a\left(  t\right)  \right]  }}J\left(  0^{+}\right)
K_{\mathrm{I}}(t)\sqrt{\frac{8}{\pi}\left[  a\left(  t\right)  - (x_{1}
-\delta a)\right]  }
\nonumber\\
+&  \int_{a}^{a+\delta a}dx_{1}\frac{K_{\mathrm{I}}(t)}{\sqrt{2\pi\left[
x_{1}-a\left(  t\right)  \right]  }}\nonumber\\
&\int_{-\infty}^{t}dt_{1}\dot
{J}\left(  t-t_{1}\right)  K_{\mathrm{I}}(t_{1})\sqrt{\frac{8}{\pi}\left[
a\left(  t_{1}\right)  -(x_{1}
-\delta a) \right]}\mathcal{H}\left[
a\left(  t_{1}\right)  -(x_{1}
-\delta a)\right]
=2\Delta\gamma\delta a
\label{equilibrium dimensional}
\end{align}
\gc{We remark that, unlike the crack-initiation criterion, in the crack propagation problem
the quantity $\delta a$ cannot be sent to zero. The reason is that, for a crack advancing
at finite speed, the singular asymptotic fields cannot remain valid all the way down to
the tip. Otherwise, the local viscoelastic dissipation density, which is of order
$\sigma_{ij}\dot{\varepsilon}_{ij}$, would become non-integrable in the vicinity of the
crack tip, leading to a divergence of the total dissipated power. This unphysical result
shows that a short-distance regularization is necessary. In the present formulation,
$\delta a$ provides such a regularization and can therefore be interpreted as a
characteristic length associated with the fracture/process zone.}
Hence, being $\delta a$ the only characteristic lenght, the time evolution of the reduced crack length $a(t)/\delta a$ must depend only on the current dimensionless stress intensity factor $K_{\mathrm{I}}\left(  t\right)  /\sqrt{2E_{0}\Delta\gamma}$ and the dimensionless creep function $E_{0}J(t)$.

\begin{figure}[ptb]
\centering\includegraphics[width=0.95\textwidth]{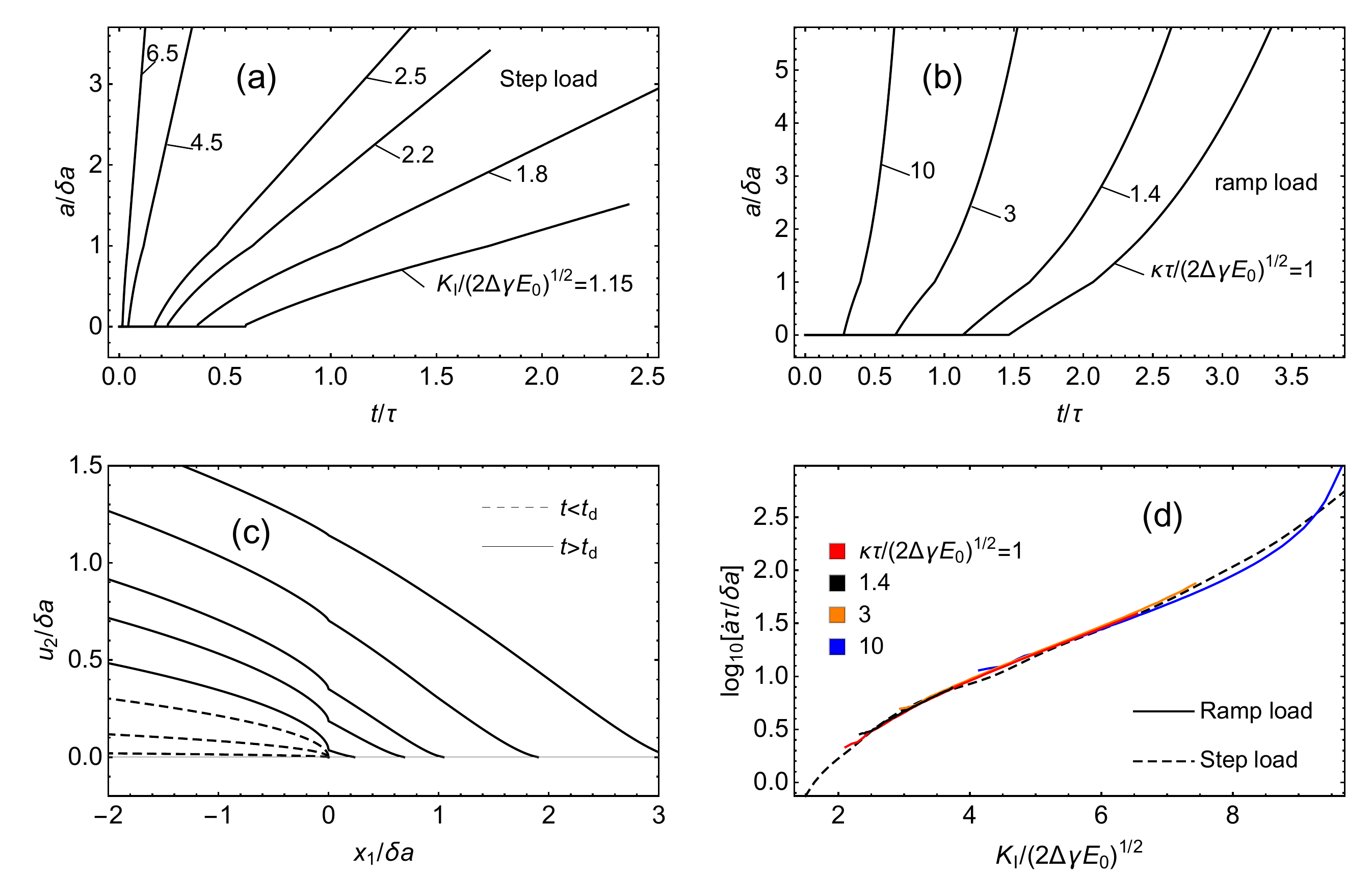}\caption{{The evolution of the dimensionless crack length $a/\delta a$ vs. the dimensionless time $t/\tau$ in response to a stress intensity factor time history $K_{\mathrm{I}}$ corresponding to (a) a step function and (b) an unbounded 
ramp. Results are shown for (a) different values of the dimensionless
stress intensity factor magnitude $K_{\mathrm{I}}/\left(2\Delta\gamma E_{0}\right)^{1/2}$ and (b) different dimensionless ramp slope $\kappa\tau/\left(2\Delta\gamma E_{0}\right)^{1/2}$ values. (c) Crack shape analysis: evolution over time of the dimensionless vertical displacement $u_2/\delta a$ along the crack in response to a step load corresponding to $K_{\mathrm{I}}\left(  t\right)
/\left(2\Delta\gamma E_{0}\right)^{1/2}=1.8\mathcal{H}\left(  t\right)  $. (d) The
instantaneous value of the dimensionless crack speed $\dot{a}\tau/\delta a$ as a function of the instantaneous dimensionless stress intensity factor $K_{\mathrm{I}}(t)/(2\Delta\gamma E_{0})^{1/2}$ for both ramp and step loading histories. All results are provided for $E_{\infty}/E_{0}=100$.}}%
\label{figPropagation}%
\end{figure}

Numerical results are reported in Fig. \ref{figPropagation}.
Specifically, Figure \ref{figPropagation}(a) and Figure \ref{figPropagation}(b) show the dimensionless crack length $a(t)/\delta a$ as a function of the dimensionless time $t/\tau$, for the step $K_{\mathrm{I}}\mathcal{H}\left(t\right)$ and unbounded ramp loading $\dot{K}_{\mathrm{I}}(t)=\kappa\mathcal{H}(t)$ cases, respectively. In the former case, we vary the value of $K_{\mathrm{I}}$, in the latter we investigate the effect of the grow rate $\kappa$.
Interestingly, Fig. \ref{figPropagation}(a) shows that, in response to a step load, the initial crack speed $\dot{a}(t_d^+)$ at the onset of crack propagation is finite and greater than zero (see the slope of the curves) and then becomes constant for $t>t_{\mathrm{s}}$, where $t_{\mathrm{s}}$ is the time at which the crack has advanced of the quantity $\delta a$.
This can be qualitatively explained observing that, along the crack axis ($x_2=0$), points at $x_{1}<0$ (i.e., in the preexisting crack) undergo creep-induced vertical displacements yet before crack propagation starts; therefore, $u_{2}(x_{1}<0,0,t<t_{\mathrm{d}})>0$, as shown by dashed lines in Fig. \ref{figPropagation} (c). 
On the other hand, points at $x_{1}>0$ do not displace before crack tip motion, and $u_{2}(x_{1}>0,0,t_{d})=0$. As a consequence, since Eq. (\ref{crack tip work balance}) enforces local equilibrium across the chacteristic length $\delta a$, at the early stage of crack propagation, with $t \lessapprox t_s$ and $a(t)\lessapprox \delta a$, the region where the crack was originally open (and creep-induced displacements already occurred) still affects the crack tip equilibrium. Once $a(t)\gtrapprox \delta a$, instead, the crack tip equilibrium mainly depends on originally unfractured material, all undergoing the same displacements time-history as the crack tip moves; as a consequence, steady-state conditions are established for $t>t_s$. 
This is clearly shown in Fig. \ref{figPropagation}(c) as the crack displacement $u_{2}$ for $0<x_{1}<a\left(t\right)$ is self-similar during crack propagation, with the typical
'trumpet shape' \cite{deGennes1996}. 

Interestingly, Eq. (\ref{equilibrium dimensional}) entails that $\dot{a}(t)/\delta a$ depends mainly on the instantaneous value of $K_{\mathrm{I}}(t)$, in agreement with previous findings \cite{Schapery1975.2,Knauss2015}, despite that Schapery and Knauss postulate different failure mechanisms. 
Calculations also confirm this feature, as shown in Fig. \ref{figPropagation}(d) comparing step and ramp loads (at different values of $\kappa$) and showing that $\dot{a}(t)/\delta a$ just depends on $K_{\mathrm{I}}$ regardless of the load time history.

\begin{figure}[ptb]
\centering\includegraphics[width=0.45\textwidth]{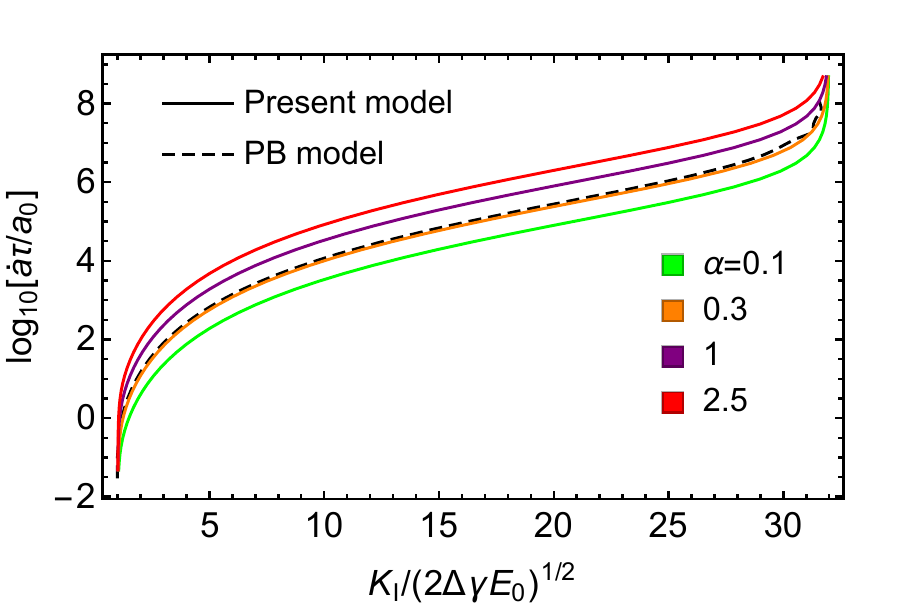}\caption{{Comparison between the present model and the Persson-Brener (PB) one: the dimensionless crack's speed $\dot{a}\tau/a_{0}$ as function of the dimensionless stress intensity factor $K_{\mathrm{I}}/(2\Delta\gamma E_{0})^{1/2}$. 
The process zone size in the limit of slow moving crack is $a_{0}=2E_{0}\Delta\gamma/\left(  2\pi\sigma_{c}^{2}\right) $ according to
PB, and $\sigma_{c}$ is a characteristic failure o yield stress. The present model results are given for process zone size estimated as $\delta a=\alpha(K_{\mathrm{I}}%
/\sqrt{2\pi}\sigma_{c})^{2}$ and different values of $\alpha$. All results are given for $E_{\infty
}/E_{0}=10^{3}$. }}%
\label{figPersson}%
\end{figure}

In light of the previous discussion, since the crack propagation speed $\dot{a}(t)$ is mainly a function of $K_{\mathrm{I}}$ and is linear with $\delta a$, we find interesting to compare our theory with existing ones postulating a load-dependent size of the process zone (i.e., $\delta a$). 
We use the Persson–Brener (PB) theory \cite{Persson2005}, assuming $\delta a=\alpha(K_{\mathrm{I}}/\sqrt{2\pi}\sigma_{c})^{2}$, where $\alpha$ is a parameter and $\sigma_{c}$ is a material characteristic failure (yielding) stress. 
The comparison is shown in Fig. \ref{figPersson}, in terms of the crack speed $\dot{a}(t)$ vs. $K_{\mathrm{I}}$; PB results refer to $\alpha=1$, while the best fit with the present theory is achieved for $\alpha\approx0.3$, which is consistent with a similar comparison \cite{Persson2021} between PB and Schapery with cohesive stress $\sigma_{0} \simeq 3\sigma_{c}$ (see also \cite{Greenwood2004}). Although PB, Schapery, and the present model differ for the description of the 'process zone', their results are in quantitative agreement (despite a minor quantitative shift) with values of $\alpha$ all order unity. Again, this suggests that the detailed description of the stress distribution within the process zone is not a key factor affecting the physics of the crack propagation, as already argued by Persson
\cite{Persson2005,Persson2021}, Greenwood \cite{Greenwood2007}, and Schapery
\cite{Schapery1975,Schapery1975.2}.

\section{Finite Element Analysis}
\begin{figure}[ptbh]
\centering\includegraphics[width=0.85\textwidth]{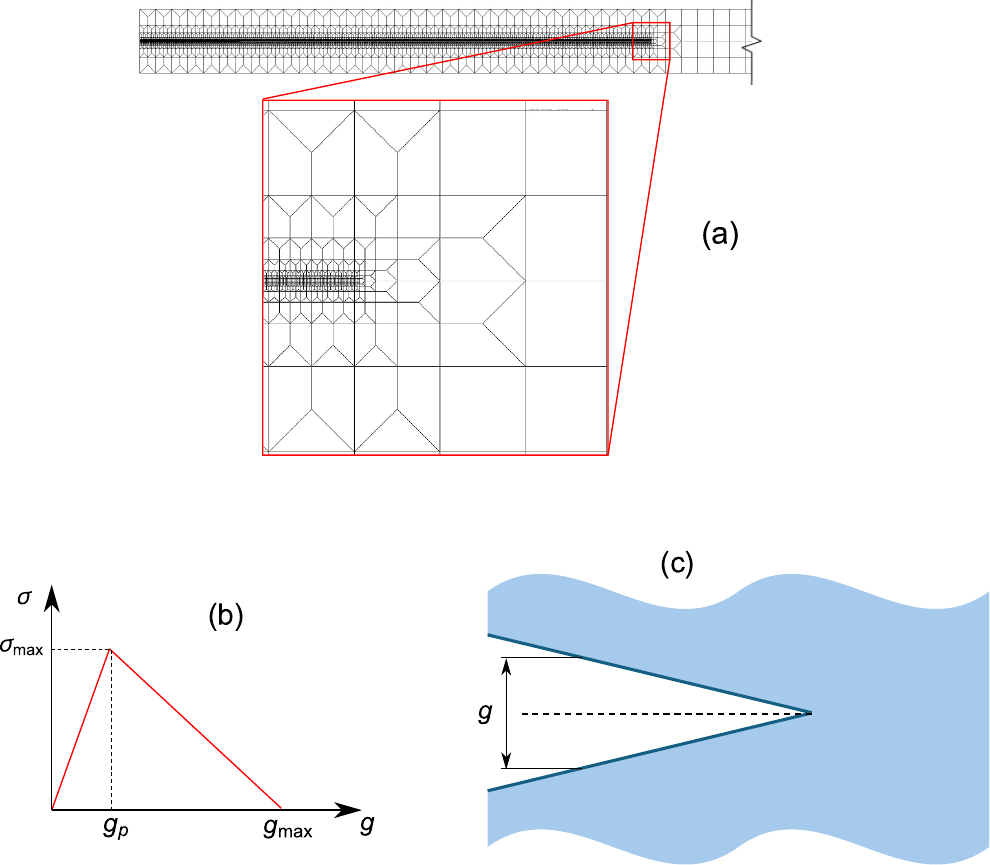}\caption{(a)
Detail of FE model with mesh coarsening. The finite-range adhesion model: (a)
the force vs gap curve; (b) the bilinear cohesive model; (c) a schematic of the crack tip.}%
\label{mesh}%
\end{figure}

In this section, we present a finite element analysis of the delay time $t_{\mathrm{d}}$ and crack length evolution $a(t>t_d)$. To
model the pure shear  specimen shown in fig.~\ref{fig1}, we consider a
rectangular domain with length $L=800~\mathrm{nm}$ and height $2h=12.8~\mathrm{nm}$, containing an pre-existing edge crack of length $a_{0}=20$~$\mathrm{nm}$, located on the left side of the specimen at mid-height.
The corresponding Finite Element model is shown in fig. \ref{mesh}, where mesh
refinement has been controlled in detail to ensure sufficient resolution for
the stress and displacements fields at the crack tip. Indeed, close to the
crack tip the mesh size is $\Delta x=0.025$~$\mathrm{nm}$, while moving far
from the high gradient region a coarser mesh is defined, ensuring
computational efficiency.

We apply the tensile force $F$ at
the midpoints of both the top and bottom boundaries of the system. This force
is introduced with an (almost) instantaneous ramp (of approximate duration
$\tau/1000$) and then remains constant for the entire simulation, so that step
load conditions can be assumed. Furthermore, multi-point constraints are
imposed to the nodes of the top and bottom boundaries, effectively
leading to a rigid motion of both edges and to a uniform tensile stress distribution $\sigma_\infty$ at the far right end of the system (see Fig. \ref{mesh}a). Quadrilateral and
triangular plane elements with linear shape functions are used to model the
2D-body in plane stress condition. The specimen is characterized by the
viscoelastic creep function provided by Eq. (\ref{creep function}). We assumed
$E_{0}=650$~$\mathrm{MPa}$, with $E_{\infty}$ $=$ $10E_{0}$, and $\tau
=10^{-3}~\mathrm{s}$. Finally, Poisson's ratio is $\nu=0.49$. To model the
interactions at the interface, a bilinear cohesive zone model is implemented
(Fig. \ref{mesh}b), characterized by three parameters: $\sigma_{\max}$, which
represents the maximum normal cohesive traction; $g_{p}$, the normal gap at
which $\sigma_{\max}$ occurs; and $g_{\max}$, the normal gap beyond which
complete de-bonding occurs. These parameters enable the description of a Mode I
dominated behavior. We assumed a surface energy $\frac{1}{2}\sigma_{\max
}g_{\max}=2\Delta\gamma=0.055$ J/m$^{2}$.

\begin{figure}[ptbh]
\centering\includegraphics[width=0.95\textwidth]{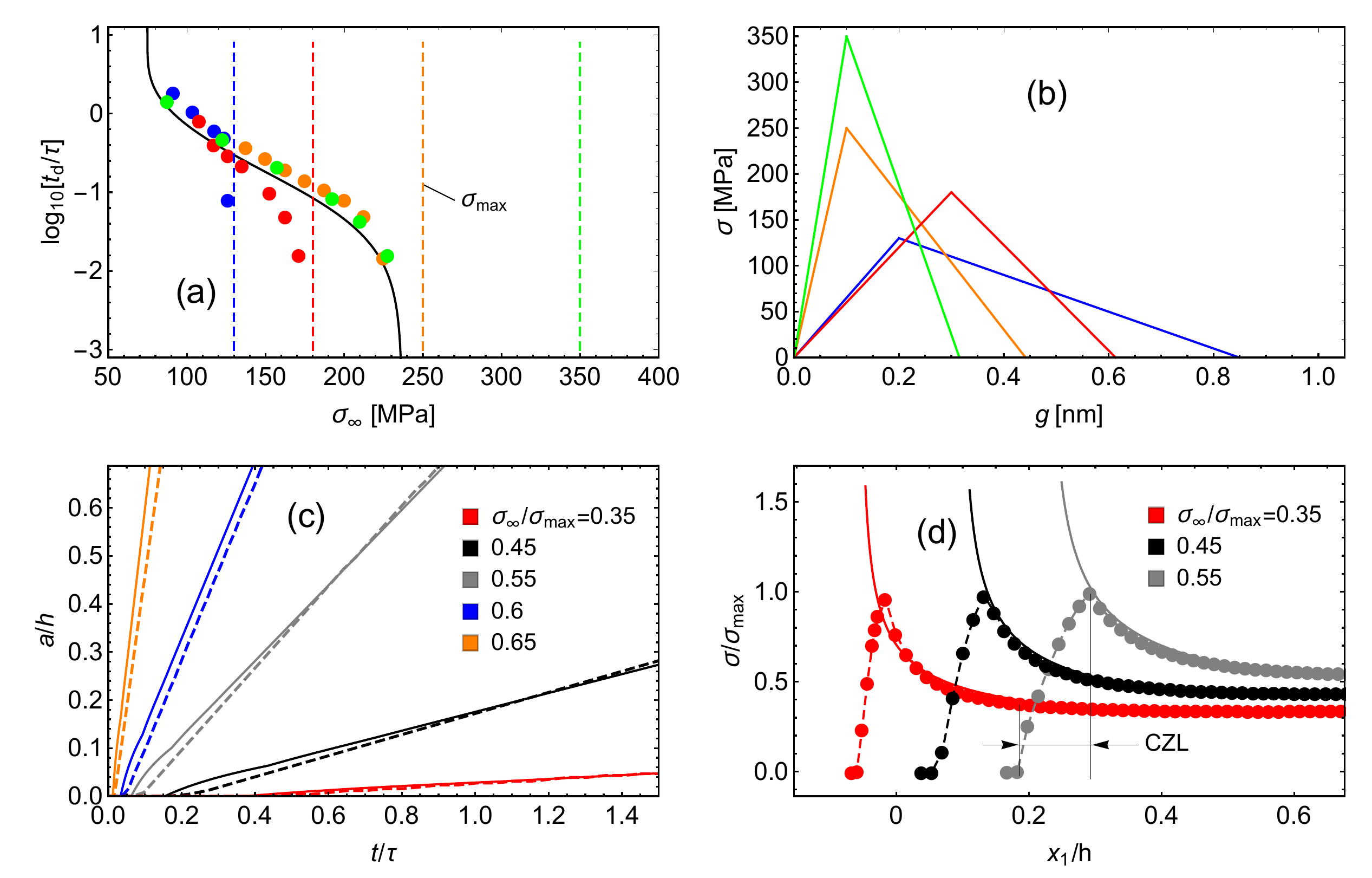}\caption{Response to step function. (a) the
dimensionless delay time $t_{\mathrm{d}}/\tau$ as a function of the remotely applied stress $\sigma_{\infty}$ as predicted by FEA (circles) and theory (solid line). FEA results are provided for the
stress vs. gap adhesion laws shown in figure (b) with corresponding colors. The maximum allowed adhesive stresses $\sigma_{\mathrm{max}}$ are reported as vertical dashed lines in (a). (c) the
dimensionless crack length $a/h$ as a function of the dimensionless time $t/\tau$ as predicted by FEA (dashed lines) and theory (solid lines). (d)
Dimensionless stress distribution $\sigma/\sigma_{\mathrm{max}}$ (FEA) close to the crack tip during crack
propagation for different dimensionless remote stress $\sigma_{\infty}/\sigma_{\mathrm{max}}$ values. Solid lines are square-root singularities with stress intensity factor $K_{\mathrm{I}}=\sigma_{\mathrm{\infty}}\sqrt{h}$. FEA results in (c) and (d) are provided for the stress-gap adhesion law reported with green line in (b).}%
\label{figFEA}%
\end{figure}

Fig. \ref{figFEA}(a) shows the dimensionless delay time in response to a step load as a function of the applied remote stress $\sigma_\infty$. Different colors refer to different bilinear cohesive parameters at the crack interface, with the corresponding law plotted in Fig. \ref{figFEA} (b). 
Numerical results are in solid agreement with the corresponding theoretical predictions provided by Eq. \ref{critical condition for crack propagaion} (black line in Fig. \ref{figFEA}a). 
More in detail, for sufficiently large values of $\sigma_{\max}\gg \sigma_\infty$ (i.e., short range of adhesion $g_{\max}$), green and orange dots in Fig. \ref{figFEA}(a) show that the gap-dependence of adhesive forces negligibly affects the delay time. This, again, confirms that the detailed description of the stress distribution at the process zone is not fundamental, in agreement with the discussion given in Sec. \ref{sec:initiation}. 
On the other hand, FEM results also show that for smaller value of $\sigma_{\max}$ (blue and red dots in Fig. \ref{figFEA}(a)), the gap-dependent adhesion does affect the results, introducing a significant deviation from theory. In fact, in these cases, fracture occurs due to quasi-uniform bond breaking \cite{Mueser2022} rather then by crack propagation, which suggests that in numerical simulations the choice of the bounded adhesion law and corresponding parameters is a critical aspect. 
Fig. \ref{figFEA}(c) shows the dimensionless crack length $a/h$ as a function of the dimensionless time. Theoretical predictions are also reported, with calculations performed according to Eq. (\ref{crack tip work balance}). Specifically, we set $\delta a$ of the same order of magnitude of the cohesive zone length ($\mathrm{CZL}$) provided by the FE results. The latter has been qualitatively defined as the lateral distance measured between the stress peak and the first traction-free point. Specifically, theoretical and FEA results present a very good overlapping for $\delta a=0.85 \mathrm{CZL}$. Dimensionless stress distributions during crack propagation are shown in Fig. \ref{figFEA}(d) for different values of the remotely applied stress $\sigma_\infty$, together with the square-root asymptotic singularity.

\gc{\section{Experimental validation of the delayed fracture theory}}

\gc{In order to further assess the descriptive capability of the proposed fracture-initiation
criterion, we carried out an additional quantitative comparison between theory and
experiments on delayed fracture in viscoelastic solids. The aim of this analysis is to
test Eq. (\ref{critical condition for crack propagaion}) against experimentally measured delay times obtained under controlled
loading histories. To this end, it is essential to have access not only to the measured
fracture-initiation times, but also to an accurate viscoelastic characterization of the
material, so that the creep function entering the theoretical formulation can be
reconstructed consistently.}

\gc{For this purpose, we made use of a previously acquired experimental dataset on cracked
PTFE specimens Ref.~\cite{Violano2023}, for which both delayed-fracture measurements and dynamic mechanical
analysis (DMA) data were available. The experimental procedures for the DMA
characterization, tensile tests, image acquisition, and crack-contour postprocessing are
described in detail in Ref.~\cite{Violano2023}. This choice was motivated by the fact
that, although many delayed-fracture experiments are available in the literature, they
often do not report the full viscoelastic input data required for a truly quantitative
comparison of the type proposed here, in particular the DMA data (or equivalent
information) needed to reconstruct the material compliance. By contrast, the PTFE dataset
considered here allows a fully consistent mechanics-based comparison between theory and
experiments.}

\gc{More specifically, the viscoelastic response of PTFE was characterized through DMA
measurements, from which the complex viscoelastic compliance was reconstructed over a
broad frequency range. The resulting response was then fitted by means of a Prony-series
representation with 60 relaxation times, so as to obtain an accurate approximation of the
creep compliance entering Eq. (\ref{critical condition for crack propagaion}). This step is crucial, since it provides the material
input required to compute the theoretical delay time under the same loading conditions
adopted in the experiments. For completeness and reproducibility, the comparison between
the experimentally reconstructed creep compliance and the fitted Prony-series
representation, together with the corresponding Prony coefficients, is reported in
\ref{app:prony}.}

\gc{The delayed-fracture experiments were performed by applying a time-varying tensile load
through a constant-speed displacement of the upper crosshead of the testing machine. The
fracture process was recorded by a camera during the entire loading history, from the
initial stretching stage up to the complete separation of the specimen into two parts. The
crack contour was determined by image postprocessing, and the position of the crack tip
was identified frame by frame. This procedure allowed us to determine the instant at
which the crack tip started to propagate, which we identify as the delayed-fracture time
$t_{\mathrm{d}}$. The loading and imaging procedures, together with the crack-tracking algorithm,
follow the methodology described in Ref.~\cite{Violano2023}.}

\gc{To compare the theoretical predictions with the experiments, the stress intensity factor
was evaluated using the geometrical correction appropriate for the finite-size cracked
specimen. Following Ref. \cite{Violano2023}, the stress intensity factor can be written as 
\begin{equation}
K_I = f(\beta)\,\sigma_\infty \sqrt{\pi a},
\label{eq:KI_finite_size}
\end{equation}
where $a$ is the initial crack length, $\beta = a/L$,  $\sigma_\infty= F/(b L)$, with $b$ and $L$ the tickness and the width of the PTFE sheet.  From  Ref. \cite{tada2000stress} we also have
\begin{equation}
f(\beta)=1.122-0.231\beta+10.55\beta^2-21.71\beta^3+30.382\beta^4.
\label{eq:shape_factor}
\end{equation}
Hence, the finite geometry affects the theoretical prediction through the corresponding
expression of $K_I$, while the energetic structure of the delayed-fracture criterion
remains unchanged.}
\begin{figure}[htbp]
    \centering 
    \includegraphics[width=0.48\textwidth]{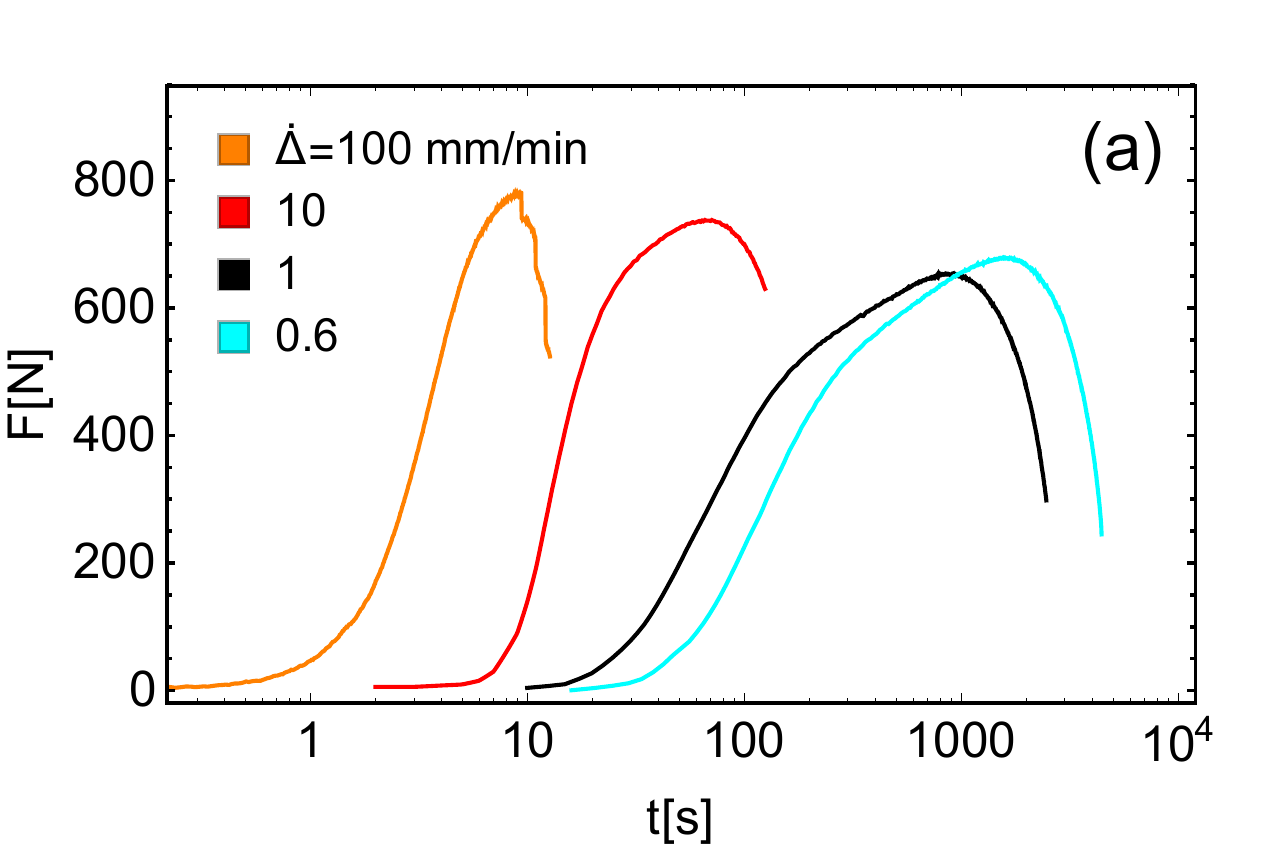}%
    \hfill
    \includegraphics[width=0.48\textwidth]{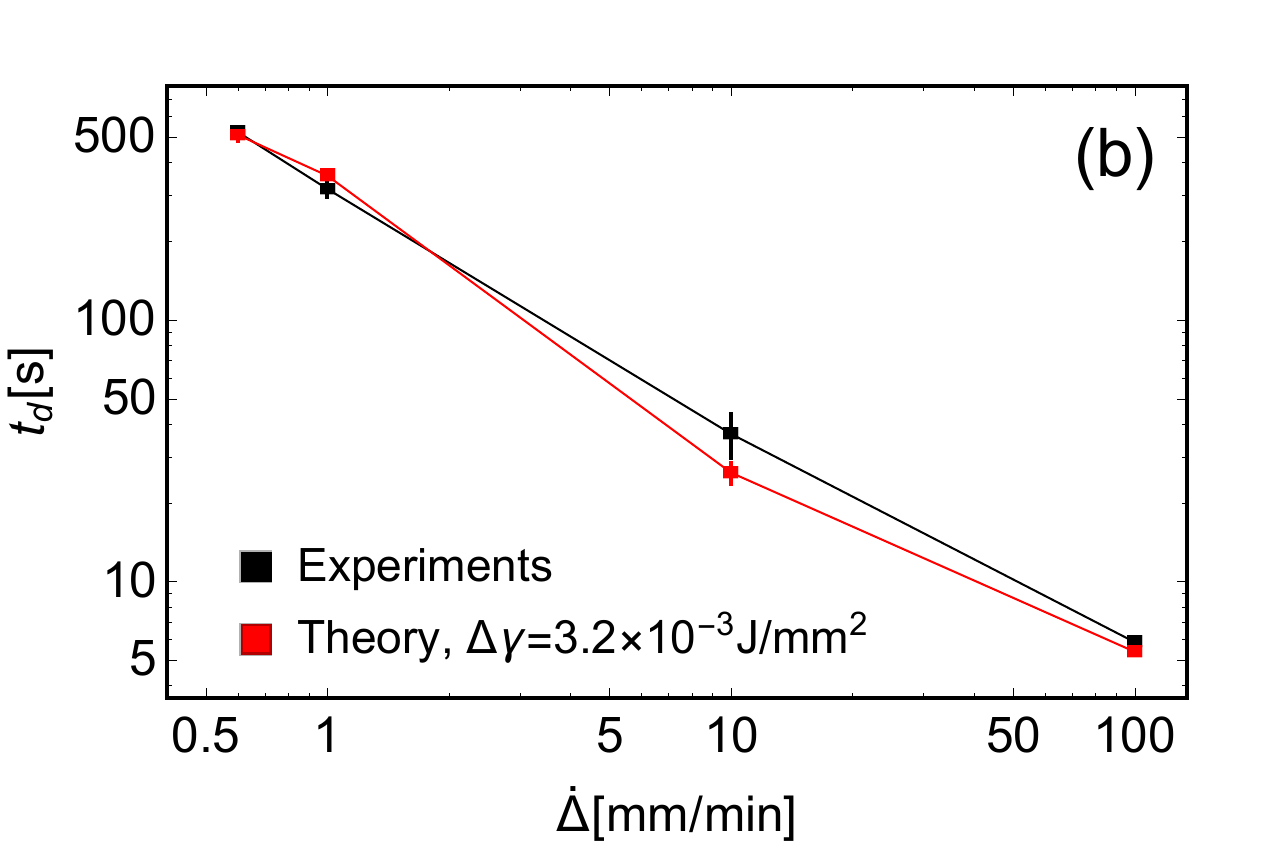}

\caption{\gc{Experimental validation of delayed fracture initiation in PTFE.
(a) Representative loading histories recorded during the tensile tests, from the initial
stretching stage up to the complete separation of the specimen into two parts.
(b) Double-logarithmic comparison between experimentally measured delay times and
theoretical predictions as functions of the crosshead speed. The theoretical results were obtained using the DMA-based viscoelastic characterization of PTFE and the corresponding Prony-series approximation of the creep compliance. The fracture energy is $\Delta\gamma = 3.2\,\mathrm{kJ\,m^{-2}}$, see Ref. \cite{Joyce2003PTFE,JoyceJoyce2004PTFE}. Very good agreement is observed over approximately two orders of magnitude in both loading time and delay time.}}
    \label{fig:delay_validation}
\end{figure}

\gc{The comparison between theory and experiments is shown in Fig.~\ref{fig:delay_validation}. Figure~\ref{fig:delay_validation}(a) reports the loading histories in a log-linear diagram from the
beginning of the tensile test up to final rupture, whereas
Fig.~\ref{fig:delay_validation}(b) shows, in double-logarithmic representation, the
experimentally measured and theoretically predicted delay times as functions of the
crosshead speed. The comparison spans approximately two orders of magnitude in both
loading time and delay time, and very good agreement is obtained throughout the full
investigated range. In particular, the experimentally measured delay times are found to
be very close to the corresponding theoretical predictions. This result provides strong
support for the ability of the proposed real-time energetic criterion to capture delayed
fracture initiation in real viscoelastic solids.}

\section{Conclusions}

We present a theory of delayed fracture in linear homogeneous and isotropic viscoelastic thin sheets. 
Moving from the Principle of Virtual Work,  we derive a closed-form rigorous expression of the available energy to propagate the crack under a generic loading time history. 

We show that, due to material viscoelasticity, the energy release rate is the sum of an elastic contribution and a non-conservative term deriving from the viscous stresses into the material. 
The critical energy condition condition for crack initiation is then expressed in terms of the time-history of the stress intensity factor modulated by the creep compliance tensor, eventually leading to a Griffith-like criterion generalized to viscoelastic solids. \gc{A key outcome of this work is, indeed, the introduction of a rigorous, path-independent $J$-integral formulation for linear viscoelastic materials, which extends the classical energy-balance framework to non-conservative solids. This provides 
a unified and general fracture criterion applicable under arbitrary loading 
histories, enabling reliable prediction of delayed fracture initiation without detailed modeling of the process zone, and directly 
exploitable in simulations and experiments.} Our analysis demonstrates that there is a load range in which the fracture is delayed depending on the specific loading history; in fact, viscoelastic sheets can sustain high (low) loads for a short (long) time before crack propagation occurs. 
Therefore, we first focus our study on the delay time for crack propagation under step, saturated ramp, and power law varying load histories.
Then we investigate the evolution of the crack over time, showing that crack propagation starts at finite speed and, once the fractured length is larger than the process zone, it reaches steady-state conditions provided that the stress intensity factor does not significantly change as the crack advances of a distance comparable with the length of the process zone. 

We successfully validate our predictions with ad hoc Finite Element simulations and experiments on delayed fracture in polymeric and soft materials. We also compare our findings with existing theories, such as those by Schapery and Persson-Brener, finding significant agreement.  We correctly predicts both the delay time and the time-dependent crack propagation by only requiring that the process zone is small compared to any relevant length of the system. This demonstrates that the stress distribution in the process zone does not play a significant role in viscoelastic crack propagation and shows that the present theory is more effective than others.

In conclusion, the presented methodology is applicable to arbitrary viscoelastic constitutive laws and loading histories, allowing one to predict crack initiation and propagation. Beyond polymers and elastomers, the approach can be extended to soft gels, composites, and biological tissues where delayed fracture is critical to performance and safety. Future works may focus on targeted experimental validation across material classes, as well as incorporating anisotropic materials and nonlinear viscoelasticity to further broaden applicability.

\section*{Acknowledgments}
The authors acknowledge support from the China-Italy International Joint Lab on Smart Tribology, from the Flagship Project 2023 – Linea B “Least Environmental impact For Sustainable mobility (LEAFS - CUP: D93C22000410001)”, Spoke 11 of the Italian National Centre for Sustainable Mobility (MOST), and from the PRIN2022 PNRR grant nr. P2022MAZHX - TRibological modellIng for sustainaBle design Of induStrial friCtiOnal inteRfacEs (TRIBOSCORE), both funded by MUR under the PNRR – NextGenerationEU programme.

\appendix
\section{J-integral formulation at the onset of crack propagation in viscoelastic materials\label{sec:app J-int}}
In this appendix, we derive the critical crack propagation condition, Eq. (\ref{critical condition for crack propagaion}), by applying the Principle of Virtual Work to a generic surface $S$ surrounding the crack tip, eventually leading to a path-independent line integral along the surface contour $\partial S$, which is commonly referred to as the $J$-integral in classical fracture mechanics \cite{Rice1968}.

Following Eq. (\ref{self-similiarity}), we rewrite the stress fields in terms of the stress intensity factor $K_I$ using the shape functions $\zeta_{ij}$ as 
\begin{equation}
    \sigma_{ij}\left(x_{1},x_{2},t\right)  =K_{\mathrm{I}}\left(  t\right)
\zeta_{ij}\left(  x_{1} - a,x_{2}\right)  \label{KI shape functions}%
\end{equation}

\begin{figure}[ptbh]
\centering\includegraphics[width=0.85\textwidth]{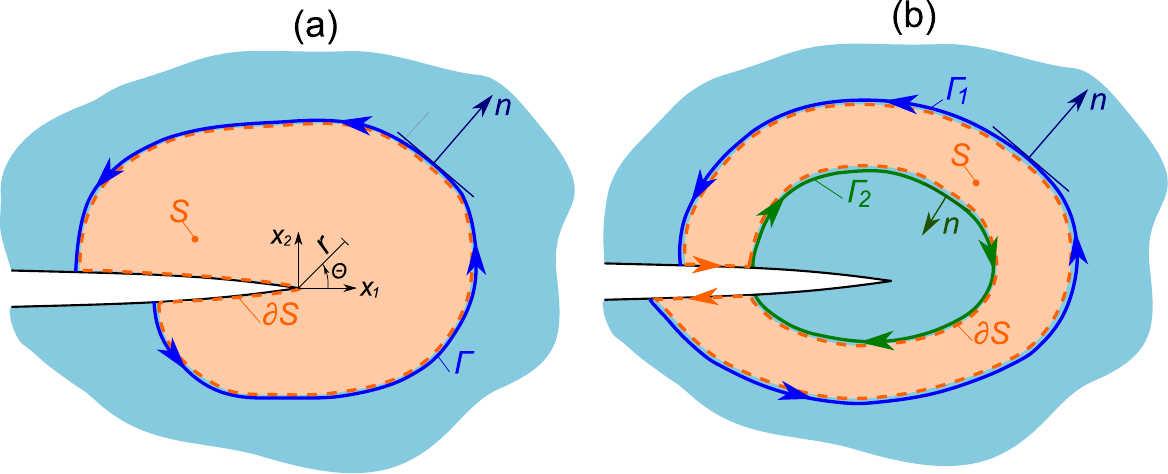}\caption{Possible integration domains $S$ including (a) and excluding (b) the crack tip. In (a), the tip centered polar coordinate system is also shown. Notably, $\Gamma$ is the portion of the domain contour $\partial S$ subjected to non-vanishing stresses.}%
\label{figJint}%
\end{figure}

Firstly, we focus on the case shown in Fig. \ref{figJint}(a), where $S$ includes the crack tip  and $\Gamma \subseteq \partial S$ is the portion of the surface contour $\partial S$ subjected to non-vanishing stresses. Using Eq. (\ref{virtual displacement}) for the virtual displacement field, the virtual work of external stresses $\delta L_{\mathrm{E}}$ given a virtual variation $\delta a$ of the crack length $a$ can be written as
\begin{equation}
          \delta L_{\mathrm{E}}=-\delta a\int_{\Gamma}ds\sigma_{ij}(x_{1},x_{2},t)n_{j}(x_{1},x_{2})\frac{\partial u_{i}(x_{1},x_{2},t)}{\partial x_{1}}-2\Delta\gamma\delta a
\label{external work J}
\end{equation}
where $\mathbf{n}$ is the local outward unit normal vector to $\Gamma$ and $ds$ is an element of arc length along $\Gamma$. 
Now, following Eqs. (\ref{interal work}, \ref{virtual strain 2}), the virtual work of internal stresses $\delta L_{\mathrm{I}}$ is
\begin{equation}
\delta L_{\mathrm{I}}=-\delta a
K_{\mathrm{I}}(t)\int_{S} d^{2}x \dot{K}_{\mathrm{I}}\left(
t\right)\conv J_{ijhk}\left(  t\right)\zeta_{ij}(x_{1}-a,x_{2})\frac
{\partial\zeta_{hk}(x_{1}-a,x_{2})}{\partial x_{1}} 
\label{surface integral}
\end{equation}
where the operator '$\conv$' denotes the commutative time convolution product.
Recalling again the major symmetry of $J_{ijhk}(t)$ (see Eq. (\ref{passaggio derivata})) and using the Green's theorem, we can write 
\begin{equation}
    \int_{S} d^{2}xJ_{ijhk}\left(  t\right)\zeta_{ij}(x_{1}-a,x_{2})\frac
{\partial\zeta_{hk}(x_{1}-a,x_{2})}{\partial x_{1}}= \int_{S} d^{2}x\frac
{\partial\phi(x_{1},x_{2},t)}{\partial x_{1}}=\int_{\Gamma} dx_{2}\phi(x_{1},x_{2},t)
\label{Green}
\end{equation}
where the line integral along $\Gamma$ is evaluated in counterclockwise sense, and 
\begin{equation}
   \phi(x_{1},x_{2},t)=\frac{1}{2}J_{ijhk}(t)\zeta_{ij}(x_{1}-a,x_{2})\zeta_{hk}(x_{1}-a,x_{2})
   \label{potential}
\end{equation}
Combining Eqs. (\ref{surface integral}-\ref{potential}), we get
\begin{equation}
\delta L_{\mathrm{I}}=-\delta a
K_{\mathrm{I}}(t)\int_{\Gamma} dx_{2}\phi(x_{1},x_{2},t)\conv \dot{K}_{\mathrm{I}}(t)
\label{internal work J}
\end{equation}

Finally, using Eqs. ( \ref{external work J}, \ref{internal work J}) the equilibrium condition $\delta L_{\mathrm{I}}=\delta L_{\mathrm{E}}$, reads as
\begin{equation}
J_{\mathrm{I}}(t)=2\Delta\gamma
\label{equilibrium J}
\end{equation}
where 
\begin{equation}
J_{\mathrm{I}}(t)=K_{\mathrm{I}}(t)\int_{\Gamma} dx_{2}\phi(x_{1},x_{2},t)\conv \dot{K}_{\mathrm{I}}(t)-\int_{\Gamma} ds\sigma_{ij}(x_{1},x_{2},t) n_{j}(x_{1},x_{2}) \frac{\partial u_{i}(x_{1},x_{2},t)}{\partial x_{1}}
\label{J integral}
\end{equation}
is the time-dependent $J$-integral for viscoelastic materials.

Importantly, although $J_{\mathrm{I}}(t)$ is calculated along a specific contour $\Gamma$, it can be shown that its value is path-independent. Indeed, considering a surface $S$ which does not include the crack tip (see Fig. \ref{figJint}(b)), there is no adhesion related virtual external work; therefore, the equilibrium condition $\delta L_{\mathrm{I}}=\delta L_{\mathrm{E}}$ rewritten in terms of Eq. (\ref{equilibrium J}) reads
\begin{equation}
J_{\mathrm{I},1}(t) = J_{\mathrm{I},2}(t)
\label{J12}
\end{equation}
where, with reference to Fig. \ref{figJint}(b), the terms $J_{\mathrm{I},1}(t)$ and $J_{\mathrm{I},2}(t)$ can be calculated using Eq. (\ref{J integral}) along the different contours $\Gamma_1$ and $\Gamma_2$, respectively, noting that $J_{\mathrm{I},2}(t)$ is evaluated in the clockwise sense and $\Gamma_2$ is the inner contour of the surface.

We remark that, although the material response is non-conservative, the quantity $J_I(t)$ defined in Eq. (\ref{J integral}) remains path-independent because it follows from the Principle of Virtual Work and the major symmetry of the creep compliance tensor. Hence, $J_I(t)$ represents the virtual work of stresses per unit virtual crack advance, rather than the release of purely elastic energy, thereby extending the $J$-integral concept to linear viscoelastic solids.
 
Moreover, according to the derivation presented in Sec. \ref{sec:initiation}, the energy release rate given by Eq. (\ref{energy release rate}) corresponds to the $J$-integral evaluated along the contour of the whole sheet. Thus, we conclude that, for any path around the crack tip we have
\begin{equation}
J_{\mathrm{I}}(t)=G(t)=K_{\mathrm{I}}\left(  t\right)  \int_{-\infty}^{t}dt_{1}J\left(
t-t_{1}\right)  \dot{K}_{\mathrm{I}}\left(  t_{1}\right)
\label{J-integral-G}
\end{equation}

As a final remark, we observe that the same expression can be derived by using asymptotic near-tip fields. In what follows, consider an integration path $\Gamma$ represented by a circle of radius $R$ centered in the crack tip, with $R$ being much smaller than any other characteristic length. The asymptotic near-tip stress shape functions are expressed in polar coordinates $(r,\theta)$ centered in the \nm{pre-existing} crack tip (see Fig. \ref{figJint}(a)); following \cite{Maugis2000} we have
\begin{align}
&\zeta_{11}(r,\theta)= \frac{1}{\sqrt{2\pi r}} \cos\left(\frac{\theta}{2}\right) \left[1 - \sin\left(\frac{\theta}{2}\right) \sin\left(\frac{3\theta}{2}\right)\right]\nonumber\\
&\zeta_{22}(r,\theta) = \frac{1}{\sqrt{2\pi r}} \cos\left(\frac{\theta}{2}\right) \left[1 + \sin\left(\frac{\theta}{2}\right) \sin\left(\frac{3\theta}{2}\right)\right] \nonumber\\
&\zeta_{12}(r,\theta) =\frac{1}{\sqrt{2\pi r}} \cos\left(\frac{\theta}{2}\right) \sin\left(\frac{\theta}{2}\right) \cos\left(\frac{3\theta}{2}\right) \label{near tip stress}
\end{align}
Moreover, we assume isotropic material with constant Poisson’s ratio $\nu$, yielding
\begin{equation}
J_{ijhk}(t) = J(t) \left[\frac{1+\nu}{2}(\delta_{ih}\delta_{jk} + \delta_{ik}\delta_{jh}) - \nu \delta_{ij}\delta_{hk}\right]
\end{equation}
Using the asymptotic fields into Eq. (\ref{potential}), and defining $\bar{\zeta}_{ij}(\theta) = \sqrt{2\pi r}\, \zeta_{ij}(r,\theta)$, we can calculate the term
\begin{equation}
\int_{\Gamma} dx_{2}\phi(x_{1},x_{2},t)=\frac{1}{4\pi} J_{ijhk}(t) \int_{0}^{2\pi} d\theta\, \cos\theta\, \bar{\zeta}_{ij}(\theta)\bar{\zeta}_{hk}(\theta)
= \frac{1-\nu}{4} J(t)
\label{first term}
\end{equation}
Focusing on the second term in Eq. (\ref{J integral}), according to the extended correspondence principle \cite{Christensen,Schapery1984,Schapery1975}, the viscoelastic displacement fields are  
\begin{equation}
u_{i}(x_{1},x_{2},t)=J(t)\conv \dot{u}_{i}^{E}(x_{1},x_{2},t)
\label{correspondence}
\end{equation}
where $u_{i}^{E}(x_{1},x_{2},t)$ are the displacement fields for an elastic material with unitary Young modulus, which can be factorized using the corresponding shape functions $\psi_{i}$. We have
\begin{equation}
u_{i}^{E}(x_{1},x_{2},t)=K_{\mathrm{I}}(t)\psi_{i}(x_{1},x_{2})
\end{equation}
and with the viscoelastic displacements derivatives given by
\begin{equation}
\frac{\partial u_{i}}{\partial x_{1}} = \frac{\partial \psi_{i}(x_{1},x_{2})}{\partial x_{1}}J(t)\conv\dot{K}_{\mathrm{I}}(t)
\label{visco disp derivatives}
\end{equation}
Close to the crack tip (see \cite{Maugis2000}), we have 
\begin{align}
&\psi_{1}(r,\theta) = (1+\nu)\sqrt{\frac{r}{2\pi}} \cos\frac{\theta}{2} \left(\frac{3-\nu}{1+\nu} - \cos\theta\right)\nonumber \\
&\psi_{2}(r,\theta) =(1+\nu)\sqrt{\frac{r}{2\pi}} \sin\frac{\theta}{2} \left(\frac{3-\nu}{1+\nu} - \cos\theta\right)
\label{elastic disp}
\end{align}
Since the integration domain is circular, using Eqs. (\ref{KI shape functions}, \ref{visco disp derivatives}) for the second term of Eq. (\ref{J integral}) we obtain
\begin{equation}
  \int_{\Gamma} ds\sigma_{ij} n_{j} \frac{\partial u_{i}}{\partial x_{1}}  =K_{\mathrm{I}}(t)R\int_{0}^{2\pi}d\theta\zeta_{ij}(R,\theta)n_{j}(\theta) \frac{\partial \psi_{i}(R,\theta)}{\partial x_{1}} J(t)\conv\dot{K}_{\mathrm{I}}(t)
\label{second term}
\end{equation}
where $n_{1}=\cos{\theta}$ and $n_{2}=\sin{\theta}$. Then, using the asymptotic fields, Eqs. (\ref{near tip stress}, \ref{elastic disp}), we obtain
\begin{equation}
R\int_{0}^{2\pi}d\theta\zeta_{ij}(R,\theta)n_{j}(\theta) \frac{\partial \psi_{i}(R,\theta)}{\partial x_{1}}=-\frac{3+\nu}{4}
\label{second term 2}
\end{equation}
Substituting Eqs. (\ref{first term}, \ref{second term}, \ref{second term 2}) into Eq. (\ref{J integral}) yields
\begin{equation}
J_{\mathrm{I}}(t)=K_{\mathrm{I}}\left(  t\right)  \int_{-\infty}^{t}dt_{1}J\left(
t-t_{1}\right)  \dot{K}_{\mathrm{I}}\left(  t_{1}\right)
\label{J integral2}
\end{equation}
which corresponds to Eq. (\ref{energy release rate}).

\section{The critical condition \nm{for fracture initiation} derived from
local energy balance\label{sec:app A}}

In this appendix, we derive the critical condition \nm{for fracture initiation (with the crack tip standing still)} Eq. (\ref{critical condition for crack propagaion}) by using the
virtual work balance already presented in a previous paper by some of the authors
\cite{Mandriota2024} and also utilized in Sec. \ref{sec:propagation} to
study crack propagation. The local energy balance is here reported for the
sake of clarity%

\begin{equation}
\frac{1}{2\delta a}\int_{a}^{a+\delta a}dx_{1}\sigma_{22}%
(x_{1},0,t)u_{2}(x_{1}-\delta a,0,t)=\Delta\gamma\label{local balance eq}%
\end{equation}
Observing that the crack has not yet propagated, the singular stress field at
the crack tip is
\begin{equation}
\sigma_{22}(x_{1},0,t)=\frac{K_{I}(t)}{\sqrt{2\pi x_{1}}}\mathcal{H}\left(
x_{1}\right)  \label{squarer2}%
\end{equation}
and the crack displacement reads
\begin{equation}
u_{2}(x_{1},0,t)=\sqrt{\frac{8}{\pi}}\mathcal{H}(-x_{1})\int_{-\infty}%
^{t}dt_{1}J\left(  t-t_{1}\right)  \dot{K}_{I}\left(  t_{1}\right)
\label{disp3}%
\end{equation}
Therefore, using Eqs. (\ref{squarer2}, \ref{disp3}) in Eq.
(\ref{local balance eq}), we get
\begin{equation}
\frac{K_{I}\left(  t\right)  }{2\delta a}\int_{0}^{\delta a}dx_{1}\sqrt
{\frac{4\left(  \delta a-x_{1}\right)  }{\pi^{2}x_{1}}}\int_{-\infty}%
^{t}dt_{1}J\left(  t-t_{1}\right)  \dot{K}_{I}\left(  t_{1}\right)
=\Delta\gamma. \label{square root work}%
\end{equation}
Recalling that
\[
\int_{0}^{\delta a}dx_{1}\sqrt{\frac{4\left(  \delta a-x_{1}\right)  }{\pi
^{2}x_{1}}}=\delta a
\]
we get%
\begin{equation}
K_{I}\left(  t\right)  \int_{-\infty}^{t}dt_{1}J\left(  t-t_{1}\right)
\dot{K}_{I}\left(  t_{1}\right)  =2\Delta\gamma.
\end{equation}
i.e. we recover Eq. (\ref{critical condition for crack propagaion}).

\section{Derivation of the major symmetry of the creep compliance
tensor.\label{sec:app B}}

Here we assume thermal equilibrium and linearity between stress and strain. We
will also make use of the microscopic time reversibility, i.e. time-reversal
invariance of Hamiltonian dynamics. Under these condition one can write a
generalized Langevin equation
\cite{Kubo1966,Mori1965,Onsager1931,Onsager1931.2} as
\begin{equation}
\int_{0}^{t}d\tau G_{ijhk}\left(  t-\tau\right)  \dot{\varepsilon}_{hk}\left(
\tau\right)  =\sigma_{ij}^{F}\left(  t\right)  \label{GLE}%
\end{equation}
Equation (\ref{GLE}) describes the thermal fluctuation of the shape of a small
material volume element $\delta V$ (about a generic point $\mathbf{x}$)
subjected to the fluctuating surface forces described by the random second
order stress tensor $\sigma_{ij}^{F}\left(  t\right)  $, which is switched on
at time $t=0$. In Eq. (\ref{GLE}) $G_{ijhk}\left(  t\right)  $ is the
relaxation fourth order tensor, which satisfies together with $J_{ijhk}\left(
t\right)  $ the causality principle and also the inversion relations%
\begin{equation}
\int d\tau G_{ijhk}\left(  t-\tau\right)  \dot{J}_{hkmn}\left(  \tau\right)
=I_{ijmn}^{\left(  4\right)  }\mathcal{H}\left(  t\right)  =\delta_{im}%
\delta_{jn}\mathcal{H}\left(  t\right)  \label{inversion rule}%
\end{equation}
where $I_{ijmn}^{\left(  4\right)  }=\delta_{im}\delta_{jn}$ is the fourth
order unit tensor. Also note that in general $G_{ijhk}\left(  t\rightarrow
+\infty\right)  =G_{ijhk}^{\infty}\neq0$. Therefore Eq. (\ref{GLE}) can be
rewritten as%
\begin{equation}
G_{ijhk}^{\infty}\varepsilon_{hk}\left(  t\right)  +\int_{0}^{t}%
d\tau\mathcal{G}_{ijhk}\left(  t-\tau\right)  \dot{\varepsilon}_{hk}\left(
\tau\right)  =\sigma_{ij}^{F}\left(  t\right)  \label{GLE2}%
\end{equation}
where we have defined $\mathcal{G}_{ijhk}\left(  t\right)  =G_{ijhk}\left(
t\right)  -G_{ijhk}^{\infty}$. Now let us multiply both hand sides of the
above equation for $\varepsilon_{nm}\left(  0\right)  $ and apply the ensemble
average linear operator $\left\langle {}\right\rangle $. We have%
\begin{equation}
G_{ijhk}^{\infty}\left\langle \varepsilon_{mn}\left(  0\right)  \varepsilon
_{hk}\left(  t\right)  \right\rangle +\int_{0}^{t}d\tau\mathcal{G}%
_{ijhk}\left(  t-\tau\right)  \left\langle \varepsilon_{mn}\left(  0\right)
\dot{\varepsilon}_{hk}\left(  \tau\right)  \right\rangle =\left\langle
\varepsilon_{mn}\left(  0\right)  \sigma_{ij}^{F}\left(  t\right)
\right\rangle \label{GLE 3}%
\end{equation}
Now, recalling that the relaxation times of the material are much longer than
the characteristic time of the fluctuating stress $\sigma_{ij}^{F}\left(
t\right)  $ we end up with the condition $\left\langle \varepsilon_{mn}\left(
0\right)  \sigma_{ij}^{F}\left(  t\right)  \right\rangle =0$, which gives
\begin{equation}
G_{ijhk}^{\infty}\left\langle \varepsilon_{mn}\left(  0\right)  \varepsilon
_{hk}\left(  t\right)  \right\rangle +\int_{0}^{t}d\tau\mathcal{G}%
_{ijhk}\left(  t-\tau\right)  \left\langle \varepsilon_{mn}\left(  0\right)
\dot{\varepsilon}_{hk}\left(  \tau\right)  \right\rangle =0 \label{GLE 4}%
\end{equation}
Note that $\left\langle \varepsilon_{mn}\left(  0\right)  \varepsilon
_{hk}\left(  t\right)  \right\rangle \ =R_{mnhk}\left(  t\right)  $ is the
strain correlation fourth order tensor. Microscopic time reversibility implies
that $\left\langle \varepsilon_{mn}\left(  0\right)  \varepsilon_{hk}\left(
t\right)  \right\rangle \ =\left\langle \varepsilon_{mn}\left(  0\right)
\varepsilon_{hk}\left(  -t\right)  \right\rangle $, and recalling that for a
stationary process $\left\langle \varepsilon_{mn}\left(  t^{\prime}\right)
\varepsilon_{hk}\left(  t^{\prime}+t\right)  \right\rangle =\left\langle
\varepsilon_{mn}\left(  0\right)  \varepsilon_{hk}\left(  t\right)
\right\rangle $ for any $t$ and $t^{\prime}$, we get%
\begin{equation}
R_{mnhk}\left(  t\right)  =\left\langle \varepsilon_{mn}\left(  0\right)
\varepsilon_{hk}\left(  t\right)  \right\rangle =\left\langle \varepsilon
_{hk}\left(  0\right)  \varepsilon_{mn}\left(  t\right)  \right\rangle
=R_{hkmn}\left(  t\right)  \label{microscopic reversibility}%
\end{equation}
Eq. (\ref{microscopic reversibility}) implies that the strain correlation
fourth order tensor possesses the major symmetry. Hence, Eq. (\ref{GLE 4}) can
be rephrased as%
\begin{equation}
G_{ijhk}^{\infty}R_{hkmn}\left(  t\right)  +\int_{0}^{t}d\tau\mathcal{G}%
_{ijhk}\left(  t-\tau\right)  \dot{R}_{hkmn}\left(  \tau\right)  =0;\qquad t>0
\label{strain correlation equations}%
\end{equation}
Now let's write $R_{hkmn}\left(  t\right)  =R_{hkmn}^{0}+X_{hkmn}\left(
t\right)  $ where $R_{hkmn}^{0}=R_{hkmn}\left(  0\right)  $, Eq.
(\ref{strain correlation equations}) becomes%
\begin{equation}
\int d\tau G_{ijhk}\left(  t-\tau\right)  \dot{X}_{hkmn}\left(  \tau\right)
=-\frac{k_{B}T}{\delta V}\delta_{im}\delta_{jn}H\left(  t\right)
\label{precursors FDT}%
\end{equation}
Where we have used the equipartition theorem
\cite{Callen1985} to write%
\[
G_{ijhk}^{\infty}R_{hkmn}^{0}=\frac{k_{B}T}{\delta V}\delta_{im}\delta_{jn}%
\]
Comparing Eq. (\ref{precursors FDT}) and \ref{inversion rule} and enforcing
linearity we get%
\begin{equation}
\dot{J}_{hkmn}\left(  t\right)  =-\frac{\delta V}{k_{B}T}\dot{X}_{hkmn}\left(
t\right)  \mathcal{H}\left(  t\right)  \label{FDT}%
\end{equation}
where $\dot{J}_{hkmn}\left(  t\right)  $ is the so-called fourth order
susceptibility tensor, Eq. (\ref{FDT}) is the fluctuation dissipation theorem
expressed in the time-domain. 
Noting that $X_{hkmn}(t)$, and thus $\dot{X}_{hkmn}(t)$, inherits the major 
symmetry of $R_{hkmn}(t)$, we also conclude that the same property is held by 
the creep compliance tensor $J_{hkmn}(t)$, thus demonstrating the statement given in Sec.\ref{sec:initiation}, i.e.%
\begin{equation}
J_{hkmn}\left(  t\right)  =J_{mnhk}\left(  t\right)
\label{strong symmetry creep compliance}%
\end{equation}
Minor symmetries $J_{ijkh}(t)=J_{jikh}(t)=J_{ijhk}(t)$ follow instead from the
symmetry of the strain and stress tensors. Also note that major symmetry plus
symmetry of strain (of stress) implies symmetry of stress (of strain).

It is worth noticing that time integration of Eq. (\ref{FDT}) gives%
\begin{equation}
J_{hkmn}\left(  t\right)  =J_{hkmn}^{0}\mathcal{H}\left(  t\right)
+\frac{\delta V}{k_{B}T}\left[  R_{hkmn}^{0}-R_{hkmn}\left(  t\right)
\right]  \mathcal{H}\left(  t\right)  \label{creep as function of correlation}%
\end{equation}
where $J_{hkmn}^{0}=J_{hkmn}\left(  0\right)  =\left(  G_{ijhk}^{0}\right)
^{-1}$, where $G_{ijhk}^{0}=G_{ijhk}\left(  0\right)  $. Recalling that
$R_{hkmn}\left(  t\rightarrow+\infty\right)  =0$ and $J_{hkmn}\left(
t\rightarrow+\infty\right)  =J_{hkmn}^{\infty}=\left(  G_{ijhk}^{\infty
}\right)  ^{-1}$ we also get%
\begin{equation}
J_{hkmn}^{\infty}-J_{hkmn}^{0}=\frac{\delta V}{k_{B}T}\left\langle
\varepsilon_{hk}\left(  0\right)  \varepsilon_{mn}\left(  0\right)
\right\rangle \label{sum rule}%
\end{equation}
which is the expression of the sum-rule for the creep compliance tensor.

\section{The role of the non-conservative internal work in the framework of PVW\label{sec:app C}}

For viscoelastic materials, the virtual work of internal stresses provided by Eq. (\ref{internal work final}) can be split as
\begin{equation}
\delta L_{I}=\delta \mathcal{E}+\delta L_{NC}
\label{C1}
\end{equation}
\begin{figure}[ptbh]
\centering\includegraphics[width=0.45\textwidth]{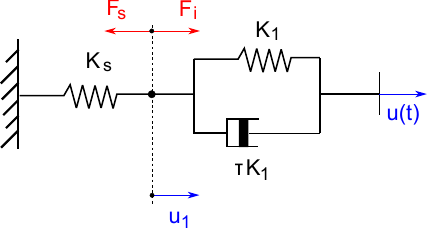}\caption{{  lumped representation of the crack propagation, red and blue arrows refer respectively to forces and displacements. }}%
\label{fig4}
\end{figure}
where $\delta L_{NC}$ is the non-conservative work contribution (vanishing in the case of purely elastic materials) and $\mathcal{E}$ is the elastic energy. Importantly, in the virtual work framework, $\delta L_{NC}$ should not be intended as a direct quantification of viscous dissipation, and it can assume both negative or positive values depending on the specific time-history of crack propagation. This conclusion can be trivially drawn by analyzing the exemplar lumped case presented in Fig. \ref{fig4}. The energy stored in the spring $K_s$ is physically equivalent to crack surface energy; therefore, the displacement $u_{1}$ can be read as a crack opening induced by a generic motion $u(t)$. Using the PVW and imposing a virtual variation $\delta u_{1}$, we have the virtual external (surface) and internal (material) work given, respectively, by 
\begin{equation}
\delta L_E=F_{s}\delta u_{1}=K_{s}u_{1}\delta u_{1}    
\end{equation}
and 
\begin{equation}
\delta L_I=F_{i}\delta u_{1}=K_{1}\left[u\left(t\right)-u_{1}\right]\delta u_{1}+\tau K_{1}\left[\dot{u}\left(t\right)-\dot{u}_{1}\right]\delta u_{1}    
\label{C3}
\end{equation}
Combining Eqs. (\ref{C1},\ref{C3}) yields
\begin{equation}
\delta L_{NC}=\tau K_{1}\left[\dot{u}\left(t\right)-\dot{u}_{1}\right]\delta u_{1}
\end{equation}
Since the PVW enforces that $\delta L_E = \delta L_I$, it is straightforward to observe that the non-conservative viscous force $\tau K_{1}\left[\dot{u}\left(t\right)-\dot{u}_{1}\right]$ can either promote or oppose crack propagation depending on the time history of the imposed displacement $u(t)$.

\gc{\section{DMA-based reconstruction of the creep compliance and Prony-series fit}\label{app:prony}}

\gc{In order to perform a quantitative comparison between the theoretical predictions of
Eq. (\ref{critical condition for crack propagaion}) and the delayed-fracture experiments, the viscoelastic creep compliance of PTFE
was reconstructed from the corresponding dynamic mechanical analysis (DMA) data.}
\begin{figure}[ptbh]
\centering\includegraphics[width=0.55\textwidth]{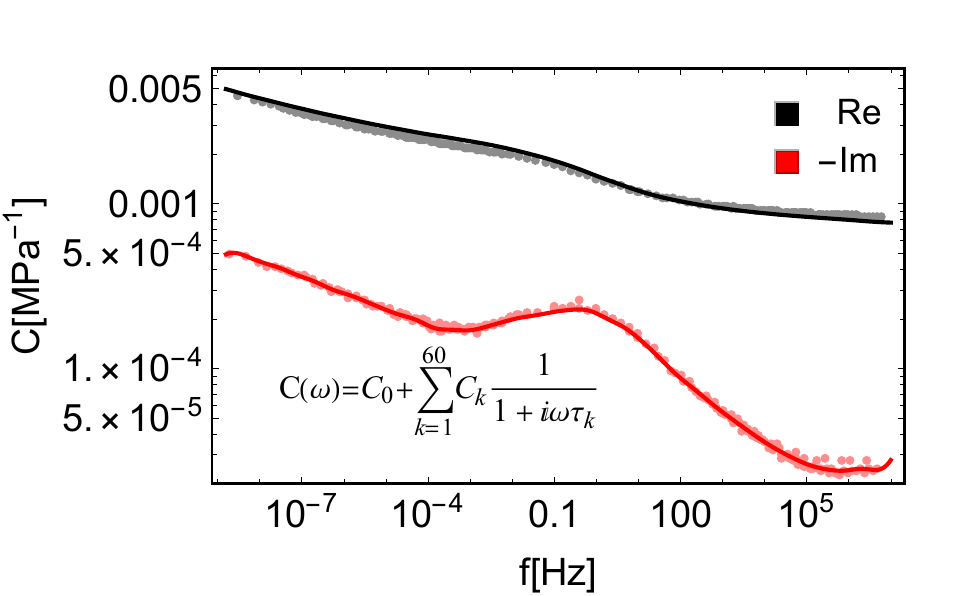}\caption{\gc{Comparison between the creep compliance reconstructed from the DMA characterization of PTFE and the corresponding Prony-series approximation used in the theoretical calculations, see also Ref. \cite{Violano2023}.}}
\label{fig:prony_coefficients}
\end{figure}

\begin{figure}[ptbh]
\centering\includegraphics[width=0.8\textwidth]{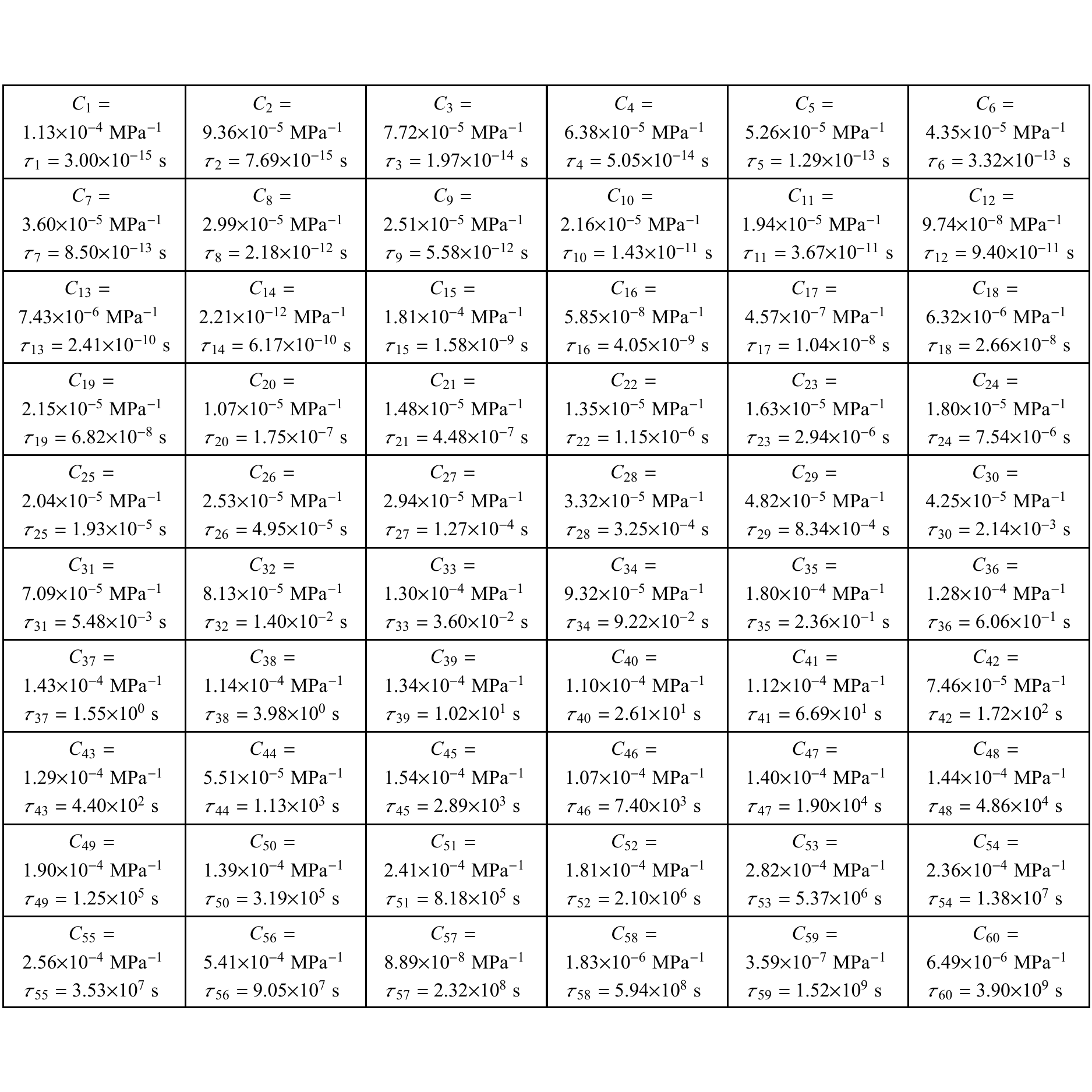}\caption{\gc{Coefficients of the Prony-series representation used to fit the creep compliance of PTFE for the delayed-fracture calculations. The high frequency compliance is $C_0 = 5.67 \times 10^{-10}\,\mathrm{MPa}^{-1}$.}}%
\label{fig:tab_prony_coefficients}%
\end{figure}

\gc{The experimental DMA characterization provides the complex viscoelastic response of the
material over a broad frequency range, from which the complex compliance can be obtained. For the purpose of the present calculations, the experimentally reconstructed compliance was approximated by means of a Prony-series representation involving a wide range of relaxation times. This approximation was used as input for the evaluation of the history-dependent convolution entering the fracture-initiation criterion. The comparison between the compliance obtained from the DMA measurements and the corresponding Prony-series fit is
shown in Fig.~\ref{fig:prony_coefficients}, while the coefficients of the Prony-series representation are reported in Fig. \ref{fig:tab_prony_coefficients} for completeness and reproducibility. The agreement between the DMA-based compliance and the Prony-series approximation is very good over the full range relevant to the delayed-fracture calculations, thus supporting the reliability of the theoretical-experimental comparison discussed in Section~5.}

\FloatBarrier

\bigskip

\bigskip

\bibliographystyle{elsarticle-harv}
\bibliography{V23_review_IJSS}

@book{Christensen,
author  = {Christensen, R. M.},
  title   = {Theory of Viscoelasticity},
  publisher = {Elsevier},
  year = {1982}
}

@article{Graham1970,
  title = {Two extending crack problems in linear viscoelasticity theory},
  volume = {27},
  number = {4},
  journal = {Quarterly of Applied Mathematics},
  publisher = {American Mathematical Society (AMS)},
  author = {Graham,  G. A. C.},
  year = {1970},
  pages = {497–507}
}

@article{Schapery1975,
  title = {A theory of crack initiation and growth in viscoelastic media: I. Theoretical development},
  volume = {11},
  number = {1},
  journal = {International Journal of Fracture},
  publisher = {Springer Science and Business Media LLC},
  author = {Schapery,  R. A.},
  year = {1975},
  pages = {141–159}
}

@article{Yeoh2003,
  title = {Fracture Mechanics of Bond Failure in the “Pure Shear” Test Piece},
  volume = {76},
  number = {2},
  journal = {Rubber Chemistry and Technology},
  publisher = {Rubber Division,  ACS},
  author = {Yeoh,  O. H.},
  year = {2003},
  month = may,
  pages = {483–494}
}

@article{Rivlin1953,
  title = {Rupture of rubber. I. Characteristic energy for tearing},
  volume = {10},
  ISSN = {1542-6238},
  url = {http://dx.doi.org/10.1002/pol.1953.120100303},
  DOI = {10.1002/pol.1953.120100303},
  number = {3},
  journal = {Journal of Polymer Science},
  publisher = {Wiley},
  author = {Rivlin,  R. S. and Thomas,  A. G.},
  year = {1953},
  month = mar,
  pages = {291–318}
}

@article{DAmico2013,
  title = {Moving cracks in viscoelastic materials: Temperature and energy-release-rate measurements},
  volume = {98},
  ISSN = {0013-7944},
  url = {http://dx.doi.org/10.1016/j.engfracmech.2012.10.026},
  DOI = {10.1016/j.engfracmech.2012.10.026},
  journal = {Engineering Fracture Mechanics},
  publisher = {Elsevier BV},
  author = {D’Amico,  F. and Carbone,  G. and Foglia,  M.M. and Galietti,  U.},
  year = {2013},
  month = jan,
  pages = {315–325}
}

@article{Griffith,
  volume = {221},
  ISSN = {2053-9258},
  url = {http://dx.doi.org/10.1098/rsta.1921.0006},
  DOI = {10.1098/rsta.1921.0006},
  number = {582–593},
  journal = {Philosophical Transactions of the Royal Society of London. Series A,  Containing Papers of a Mathematical or Physical Character},
  publisher = {The Royal Society},
author={Griffith, A.A.},
  year = {1921},
  month = jan,
  pages = {163–198}
}

@article{Shrimali2023,
  title = {The “pure-shear” fracture test for viscoelastic elastomers and its revelation on Griffith fracture},
  volume = {58},
  ISSN = {2352-4316},
  url = {http://dx.doi.org/10.1016/j.eml.2022.101944},
  DOI = {10.1016/j.eml.2022.101944},
  journal = {Extreme Mechanics Letters},
  publisher = {Elsevier BV},
  author = {Shrimali,  Bhavesh and Lopez-Pamies,  Oscar},
  year = {2023},
  month = jan,
  pages = {101944}
}

@article{Shrimali2023.2,
  title = {The delayed fracture test for viscoelastic elastomers},
  volume = {242},
  ISSN = {1573-2673},
  url = {http://dx.doi.org/10.1007/s10704-023-00700-3},
  DOI = {10.1007/s10704-023-00700-3},
  number = {1},
  journal = {International Journal of Fracture},
  publisher = {Springer Science and Business Media LLC},
  author = {Shrimali,  B. and Lopez-Pamies,  O.},
  year = {2023},
  month = jun,
  pages = {23–38}
}

@article{Mandriota2024,
  title = {Enhancement of adhesion strength in viscoelastic unsteady contacts},
  volume = {192},
  ISSN = {0022-5096},
  url = {http://dx.doi.org/10.1016/j.jmps.2024.105826},
  DOI = {10.1016/j.jmps.2024.105826},
  journal = {Journal of the Mechanics and Physics of Solids},
  publisher = {Elsevier BV},
  author = {Mandriota,  C. and Menga,  N. and Carbone,  G.},
  year = {2024},
  month = nov,
  pages = {105826}
}

@article{Mueser2022,
  title = {Crack and pull-off dynamics of adhesive,  viscoelastic solids},
  volume = {137},
  ISSN = {1286-4854},
  url = {http://dx.doi.org/10.1209/0295-5075/ac535c},
  DOI = {10.1209/0295-5075/ac535c},
  number = {3},
  journal = {Europhysics Letters},
  publisher = {IOP Publishing},
  author = {M\"{u}ser,  Martin H. and Persson,  Bo N. J.},
  year = {2022},
  month = feb,
  pages = {36004}
}

@article{Tang2017,
  title = {Fatigue fracture of hydrogels},
  volume = {10},
  ISSN = {2352-4316},
  url = {http://dx.doi.org/10.1016/j.eml.2016.09.010},
  DOI = {10.1016/j.eml.2016.09.010},
  journal = {Extreme Mechanics Letters},
  publisher = {Elsevier BV},
  author = {Tang,  Jingda and Li,  Jianyu and Vlassak,  Joost J. and Suo,  Zhigang},
  year = {2017},
  month = jan,
  pages = {24–31}
}

@article{Carbone2022,
  title = {Theory of viscoelastic adhesion and friction},
  volume = {56},
  ISSN = {2352-4316},
  url = {http://dx.doi.org/10.1016/j.eml.2022.101877},
  DOI = {10.1016/j.eml.2022.101877},
  journal = {Extreme Mechanics Letters},
  publisher = {Elsevier BV},
  author = {Carbone,  G. and Mandriota,  C. and Menga,  N.},
  year = {2022},
  month = oct,
  pages = {101877}
}

@article{Mandriota2024.2,
  title = {Adhesive contact mechanics of viscoelastic materials},
  volume = {290},
  ISSN = {0020-7683},
  url = {http://dx.doi.org/10.1016/j.ijsolstr.2024.112685},
  DOI = {10.1016/j.ijsolstr.2024.112685},
  journal = {International Journal of Solids and Structures},
  publisher = {Elsevier BV},
  author = {Mandriota,  C. and Menga,  N. and Carbone,  G.},
  year = {2024},
  month = mar,
  pages = {112685}
}

@article{Charmet1996,
  title = {Adhesive contact and rolling of a rigid cylinder under the pull of gravity on the underside of a smooth-surfaced sheet of rubber},
  volume = {16},
  ISSN = {0143-7496},
  url = {http://dx.doi.org/10.1016/S0143-7496(96)00013-9},
  DOI = {10.1016/s0143-7496(96)00013-9},
  number = {4},
  journal = {International Journal of Adhesion and Adhesives},
  publisher = {Elsevier BV},
  author = {Charmet,  J.-C. and Barquins,  M.},
  year = {1996},
  month = jan,
  pages = {249–254}
}

@article{Afferrante2022,
  title = {On the effective surface energy in viscoelastic Hertzian contacts},
  volume = {158},
  ISSN = {0022-5096},
  url = {http://dx.doi.org/10.1016/j.jmps.2021.104669},
  DOI = {10.1016/j.jmps.2021.104669},
  journal = {Journal of the Mechanics and Physics of Solids},
  publisher = {Elsevier BV},
  author = {Afferrante,  L. and Violano,  G.},
  year = {2022},
  month = jan,
  pages = {104669}
}

@article{Lorenz2013,
  title = {Adhesion: role of bulk viscoelasticity and surface roughness},
  volume = {25},
  ISSN = {1361-648X},
  url = {http://dx.doi.org/10.1088/0953-8984/25/22/225004},
  DOI = {10.1088/0953-8984/25/22/225004},
  number = {22},
  journal = {Journal of Physics: Condensed Matter},
  publisher = {IOP Publishing},
  author = {Lorenz,  B and Krick,  B A and Mulakaluri,  N and Smolyakova,  M and Dieluweit,  S and Sawyer,  W G and Persson,  B N J},
  year = {2013},
  month = may,
  pages = {225004}
}

@article{Bonn1998,
  title = {Delayed Fracture of an Inhomogeneous Soft Solid},
  volume = {280},
  ISSN = {1095-9203},
  url = {http://dx.doi.org/10.1126/science.280.5361.265},
  DOI = {10.1126/science.280.5361.265},
  number = {5361},
  journal = {Science},
  publisher = {American Association for the Advancement of Science (AAAS)},
  author = {Bonn,  Daniel and Kellay,  Hamid and Prochnow,  Michaël and Ben-Djemiaa,  Karim and Meunier,  Jacques},
  year = {1998},
  month = apr,
  pages = {265–267}
}

@article{Skrzeszewska2010,
  title = {Fracture and Self-Healing in a Well-Defined Self-Assembled Polymer Network},
  volume = {43},
  ISSN = {1520-5835},
  url = {http://dx.doi.org/10.1021/ma1000173},
  DOI = {10.1021/ma1000173},
  number = {7},
  journal = {Macromolecules},
  publisher = {American Chemical Society (ACS)},
  author = {Skrzeszewska,  Paulina J. and Sprakel,  Joris and de Wolf,  Frits A. and Fokkink,  Remco and Cohen Stuart,  Martien A. and van der Gucht,  Jasper},
  year = {2010},
  month = mar,
  pages = {3542–3548}
}

@article{Karobi2016,
  title = {Creep Behavior and Delayed Fracture of Tough Polyampholyte Hydrogels by Tensile Test},
  volume = {49},
  ISSN = {1520-5835},
  url = {http://dx.doi.org/10.1021/acs.macromol.6b01016},
  DOI = {10.1021/acs.macromol.6b01016},
  number = {15},
  journal = {Macromolecules},
  publisher = {American Chemical Society (ACS)},
  author = {Karobi,  Sadia Nazneen and Sun,  Tao Lin and Kurokawa,  Takayuki and Luo,  Feng and Nakajima,  Tasuku and Nonoyama,  Takayuki and Gong,  Jian Ping},
  year = {2016},
  month = jul,
  pages = {5630–5636}
}

@article{Brenner2013,
  title = {Failure in a soft gel: Delayed failure and the dynamic yield stress},
  volume = {196},
  ISSN = {0377-0257},
  url = {http://dx.doi.org/10.1016/j.jnnfm.2012.12.011},
  DOI = {10.1016/j.jnnfm.2012.12.011},
  journal = {Journal of Non-Newtonian Fluid Mechanics},
  publisher = {Elsevier BV},
  author = {Brenner,  Tom and Matsukawa,  Shingo and Nishinari,  Katsuyoshi and Johannsson,  Ragnar},
  year = {2013},
  month = jun,
  pages = {1–7}
}

@article{Lindstrm2012,
  title = {Structures,  stresses,  and fluctuations in the delayed failure of colloidal gels},
  volume = {8},
  ISSN = {1744-6848},
  url = {http://dx.doi.org/10.1039/C2SM06723D},
  DOI = {10.1039/c2sm06723d},
  number = {13},
  journal = {Soft Matter},
  publisher = {Royal Society of Chemistry (RSC)},
  author = {Lindstr\"{o}m,  Stefan B. and Kodger,  Thomas E. and Sprakel,  Joris and Weitz,  David A.},
  year = {2012},
  pages = {3657}
}

@article{Persson2005,
  title = {Crack propagation in viscoelastic solids},
  volume = {71},
  ISSN = {1550-2376},
  url = {http://dx.doi.org/10.1103/PhysRevE.71.036123},
  DOI = {10.1103/physreve.71.036123},
  number = {3},
  journal = {Physical Review E},
  publisher = {American Physical Society (APS)},
  author = {Persson,  B. N. J. and Brener,  E. A.},
  year = {2005},
  month = mar 
}

@article{Mishra2018,
  title = {Investigation of failure behavior of a thermoplastic elastomer gel},
  volume = {14},
  ISSN = {1744-6848},
  url = {http://dx.doi.org/10.1039/c8sm01397g},
  DOI = {10.1039/c8sm01397g},
  number = {39},
  journal = {Soft Matter},
  publisher = {Royal Society of Chemistry (RSC)},
  author = {Mishra,  Satish and Badani Prado,  Rosa Maria and Lacy,  Thomas E. and Kundu,  Santanu},
  year = {2018},
  pages = {7958–7969}
}

@article{Knauss2015,
  title = {A review of fracture in viscoelastic materials},
  volume = {196},
  ISSN = {1573-2673},
  url = {http://dx.doi.org/10.1007/s10704-015-0058-6},
  DOI = {10.1007/s10704-015-0058-6},
  number = {1–2},
  journal = {International Journal of Fracture},
  publisher = {Springer Science and Business Media LLC},
  author = {Knauss,  Wolfgang G.},
  year = {2015},
  month = nov,
  pages = {99–146}
}

@article{Ciavarella2021,
  title = {A comparison of crack propagation theories in viscoelastic materials},
  volume = {116},
  ISSN = {0167-8442},
  url = {http://dx.doi.org/10.1016/j.tafmec.2021.103113},
  DOI = {10.1016/j.tafmec.2021.103113},
  journal = {Theoretical and Applied Fracture Mechanics},
  publisher = {Elsevier BV},
  author = {Ciavarella,  Michele and Cricri,  G. and McMeeking,  R.},
  year = {2021},
  month = dec,
  pages = {103113}
}

@article{Schapery1975.2,
  title = {A theory of crack initiation and growth in viscoelastic media II. Approximate methods of analysis},
  volume = {11},
  ISSN = {1573-2673},
  url = {http://dx.doi.org/10.1007/BF00033526},
  DOI = {10.1007/bf00033526},
  number = {3},
  journal = {International Journal of Fracture},
  publisher = {Springer Science and Business Media LLC},
  author = {Schapery,  R. A.},
  year = {1975},
  month = jun,
  pages = {369–388}
}

@article{Schapery2022,
  title = {A theory of viscoelastic crack growth: revisited},
  volume = {233},
  ISSN = {1573-2673},
  url = {http://dx.doi.org/10.1007/s10704-021-00605-z},
  DOI = {10.1007/s10704-021-00605-z},
  number = {1},
  journal = {International Journal of Fracture},
  publisher = {Springer Science and Business Media LLC},
  author = {Schapery,  R. A.},
  year = {2022},
  month = jan,
  pages = {1–16}
}

@article{Frankiewicz1972,
  title = {Criteria for delayed fracture in solids and their experimental verification},
  volume = {4},
  ISSN = {0013-7944},
  url = {http://dx.doi.org/10.1016/0013-7944(72)90040-9},
  DOI = {10.1016/0013-7944(72)90040-9},
  number = {2},
  journal = {Engineering Fracture Mechanics},
  publisher = {Elsevier BV},
  author = {Frankiewicz,  K. and Stankowski,  S. and Wnuk,  M.P.},
  year = {1972},
  month = jun,
  pages = {245–266}
}

@article{Wnuk1970,
  title = {Delayed fracture in viscoelastic-plastic solids},
  volume = {6},
  ISSN = {0020-7683},
  url = {http://dx.doi.org/10.1016/0020-7683(70)90009-0},
  DOI = {10.1016/0020-7683(70)90009-0},
  number = {7},
  journal = {International Journal of Solids and Structures},
  publisher = {Elsevier BV},
  author = {Wnuk,  Milosz P. and Knauss,  Wolfgang G.},
  year = {1970},
  month = jul,
  pages = {995–1009}
}

@article{Williams1965,
  title = {Initiation and Growth of Viscoelastic Fracture},
  volume = {1},
  ISSN = {1573-2673},
  url = {http://dx.doi.org/10.1007/BF03545561},
  DOI = {10.1007/bf03545561},
  number = {4},
  journal = {International journal of fracture mechanics},
  publisher = {Springer Science and Business Media LLC},
  author = {Williams,  M. L.},
  year = {1965},
  month = dec,
  pages = {292–310}
}

@article{Knauss1970,
  title = {Delayed failure — the Griffith problem for linearly viscoelastic materials},
  volume = {6},
  ISSN = {1573-2673},
  url = {http://dx.doi.org/10.1007/BF00183655},
  DOI = {10.1007/bf00183655},
  number = {1},
  journal = {International Journal of Fracture Mechanics},
  publisher = {Springer Science and Business Media LLC},
  author = {Knauss,  W. G.},
  year = {1970},
  month = mar,
  pages = {7–20}
}

@article{Persson2021,
  title = {On Opening Crack Propagation in Viscoelastic Solids},
  volume = {69},
  ISSN = {1573-2711},
  url = {http://dx.doi.org/10.1007/s11249-021-01494-y},
  DOI = {10.1007/s11249-021-01494-y},
  number = {3},
  journal = {Tribology Letters},
  publisher = {Springer Science and Business Media LLC},
  author = {Persson,  B. N. J.},
  year = {2021},
  month = aug 
}

@article{deGennes1996,
  title = {Soft Adhesives},
  volume = {12},
  ISSN = {1520-5827},
  url = {http://dx.doi.org/10.1021/la950886y},
  DOI = {10.1021/la950886y},
  number = {19},
  journal = {Langmuir},
  publisher = {American Chemical Society (ACS)},
  author = {de Gennes,  P. G.},
  year = {1996},
  month = jan,
  pages = {4497–4500}
}

@article{Greenwood2004,
  title = {The theory of viscoelastic crack propagation and healing},
  volume = {37},
  ISSN = {1361-6463},
  url = {http://dx.doi.org/10.1088/0022-3727/37/18/011},
  DOI = {10.1088/0022-3727/37/18/011},
  number = {18},
  journal = {Journal of Physics D: Applied Physics},
  publisher = {IOP Publishing},
  author = {Greenwood,  J A},
  year = {2004},
  month = sep,
  pages = {2557–2569}
}

@article{Greenwood2007,
  title = {Viscoelastic crack propagation and closing with Lennard-Jones surface forces},
  volume = {40},
  ISSN = {1361-6463},
  url = {http://dx.doi.org/10.1088/0022-3727/40/6/025},
  DOI = {10.1088/0022-3727/40/6/025},
  number = {6},
  journal = {Journal of Physics D: Applied Physics},
  publisher = {IOP Publishing},
  author = {Greenwood,  J A},
  year = {2007},
  month = mar,
  pages = {1769–1777}
}

@article{Violano2023,
  title = {Crack propagation in viscoelastic finite-sized solids: theory and experiments},
  volume = {1275},
  ISSN = {1757-899X},
  url = {http://dx.doi.org/10.1088/1757-899X/1275/1/012043},
  DOI = {10.1088/1757-899x/1275/1/012043},
  number = {1},
  journal = {IOP Conference Series: Materials Science and Engineering},
  publisher = {IOP Publishing},
  author = {Violano,  G and Carolis,  S De and Palmieri,  M E and Carbone,  G and Tricarico,  L and Demelio,  G P and Afferrante,  L},
  year = {2023},
  month = feb,
  pages = {012043}
}

@article{Awaja2016,
  title = {Cracks,  microcracks and fracture in polymer structures: Formation,  detection,  autonomic repair},
  volume = {83},
  ISSN = {0079-6425},
  url = {http://dx.doi.org/10.1016/j.pmatsci.2016.07.007},
  DOI = {10.1016/j.pmatsci.2016.07.007},
  journal = {Progress in Materials Science},
  publisher = {Elsevier BV},
  author = {Awaja,  Firas and Zhang,  Shengnan and Tripathi,  Manoj and Nikiforov,  Anton and Pugno,  Nicola},
  year = {2016},
  month = oct,
  pages = {536–573}
}

@article{Choi2005,
  title = {Fracture initiation associated with chemical degradation: observation and modeling},
  volume = {42},
  ISSN = {0020-7683},
  url = {http://dx.doi.org/10.1016/j.ijsolstr.2004.06.028},
  DOI = {10.1016/j.ijsolstr.2004.06.028},
  number = {2},
  journal = {International Journal of Solids and Structures},
  publisher = {Elsevier BV},
  author = {Choi,  Byoung-Ho and Zhou,  Zhenwen and Chudnovsky,  Alexander and Stivala,  Salvatore S. and Sehanobish,  Kalyan and Bosnyak,  Clive P.},
  year = {2005},
  month = jan,
  pages = {681–695}
}

@article{Irez2020,
  title = {Fracture Toughness Analysis of Epoxy-Recycled Rubber-Based Composite Reinforced with Graphene Nanoplatelets for Structural Applications in Automotive and Aeronautics},
  volume = {12},
  ISSN = {2073-4360},
  url = {http://dx.doi.org/10.3390/polym12020448},
  DOI = {10.3390/polym12020448},
  number = {2},
  journal = {Polymers},
  publisher = {MDPI AG},
  author = {Irez,  Alaeddin Burak and Bayraktar,  Emin and Miskioglu,  Ibrahim},
  year = {2020},
  month = feb,
  pages = {448}
}

@article{Kuduzovi2014,
  title = {Investigations into the delayed fracture susceptibility of 34CrNiMo6 steel,  and the opportunities for its application in ultra-high-strength bolts and fasteners},
  volume = {590},
  ISSN = {0921-5093},
  url = {http://dx.doi.org/10.1016/j.msea.2013.10.019},
  DOI = {10.1016/j.msea.2013.10.019},
  journal = {Materials Science and Engineering: A},
  publisher = {Elsevier BV},
  author = {Kuduzović,  A. and Poletti,  M.C and Sommitsch,  C. and Domankova,  M. and Mitsche,  S. and Kienreich,  R.},
  year = {2014},
  month = jan,
  pages = {66–73}
}

@article{Majumder2022,
  title = {Nanomechanical testing in drug delivery: Theory,  applications,  and emerging trends},
  volume = {183},
  ISSN = {0169-409X},
  url = {http://dx.doi.org/10.1016/j.addr.2022.114167},
  DOI = {10.1016/j.addr.2022.114167},
  journal = {Advanced Drug Delivery Reviews},
  publisher = {Elsevier BV},
  author = {Majumder,  Sushmita and Sun,  Changquan Calvin and Mara,  Nathan A.},
  year = {2022},
  month = apr,
  pages = {114167}
}

@article{Sree2023,
  title = {Damage and Fracture Mechanics of Porcine Subcutaneous Tissue Under Tensile Loading},
  volume = {51},
  ISSN = {1573-9686},
  url = {http://dx.doi.org/10.1007/s10439-023-03233-x},
  DOI = {10.1007/s10439-023-03233-x},
  number = {9},
  journal = {Annals of Biomedical Engineering},
  publisher = {Springer Science and Business Media LLC},
  author = {Sree,  Vivek D. and Toaquiza-Tubon,  John D. and Payne,  Jordanna and Solorio,  Luis and Tepole,  Adrian Buganza},
  year = {2023},
  month = may,
  pages = {2056–2069}
}

@article{Wang2016,
  title = {Stretch‐Induced Drug Delivery from Superhydrophobic Polymer Composites: Use of Crack Propagation Failure Modes for Controlling Release Rates},
  volume = {55},
  ISSN = {1521-3773},
  url = {http://dx.doi.org/10.1002/anie.201511052},
  DOI = {10.1002/anie.201511052},
  number = {8},
  journal = {Angewandte Chemie International Edition},
  publisher = {Wiley},
  author = {Wang,  Julia and Kaplan,  Jonah A. and Colson,  Yolonda L. and Grinstaff,  Mark W.},
  year = {2016},
  month = jan,
  pages = {2796–2800}
}

@article{TabiloMunizaga2005,
  title = {Rheology for the food industry},
  volume = {67},
  ISSN = {0260-8774},
  url = {http://dx.doi.org/10.1016/j.jfoodeng.2004.05.062},
  DOI = {10.1016/j.jfoodeng.2004.05.062},
  number = {1–2},
  journal = {Journal of Food Engineering},
  publisher = {Elsevier BV},
  author = {Tabilo-Munizaga,  Gipsy and Barbosa-Cánovas,  Gustavo V.},
  year = {2005},
  month = mar,
  pages = {147–156}
}

@inbook{BarbosaCnovas1996,
  title = {The Rheology of Semiliquid Foods},
  ISBN = {9780120164394},
  ISSN = {1043-4526},
  url = {http://dx.doi.org/10.1016/S1043-4526(08)60073-X},
  DOI = {10.1016/s1043-4526(08)60073-x},
  booktitle = {Advances in Food and Nutrition Research},
  publisher = {Elsevier},
  author = {Barbosa-Cánovas,  Gustavo V. and Kokini,  Jozef L. and Ma,  Li and Ibarz,  Albert},
  year = {1996},
  pages = {1–69}
}

@inproceedings{Viano1986,
  series = {GMDMEETING},
  title = {Biomechanics of Bone and Tissue: A Review of Material Properties and Failure Characteristics},
  ISSN = {0148-7191},
  url = {http://dx.doi.org/10.4271/861923},
  DOI = {10.4271/861923},
  booktitle = {SAE Technical Paper Series},
  publisher = {SAE International},
  author = {Viano,  David C.},
  year = {1986},
  month = oct,
  collection = {GMDMEETING}
}

@article{Zioupos1998,
  title = {Recent developments in the study of failure of solid biomaterials and bone: ‘fracture’ and ‘pre-fracture’ toughness},
  volume = {6},
  ISSN = {0928-4931},
  url = {http://dx.doi.org/10.1016/S0928-4931(98)00033-2},
  DOI = {10.1016/s0928-4931(98)00033-2},
  number = {1},
  journal = {Materials Science and Engineering: C},
  publisher = {Elsevier BV},
  author = {Zioupos,  P.},
  year = {1998},
  month = sep,
  pages = {33–40}
}

@article{Ateshian2022,
  title = {Damage Mechanics of Biological Tissues in Relation to Viscoelasticity},
  ISSN = {1528-8951},
  url = {http://dx.doi.org/10.1115/1.4056063},
  DOI = {10.1115/1.4056063},
  journal = {Journal of Biomechanical Engineering},
  publisher = {ASME International},
  author = {Ateshian,  Gerard A. and Kroupa,  Kimberly and Petersen,  Courtney A. and Zimmerman,  Brandon and Maas,  Steve A. and Weiss,  Jeffrey A.},
  year = {2022},
  month = oct 
}

@article{Harewood2007,
  title = {Modeling of Size Dependent Failure in Cardiovascular Stent Struts under Tension and Bending},
  volume = {35},
  ISSN = {1573-9686},
  url = {http://dx.doi.org/10.1007/s10439-007-9326-6},
  DOI = {10.1007/s10439-007-9326-6},
  number = {9},
  journal = {Annals of Biomedical Engineering},
  publisher = {Springer Science and Business Media LLC},
  author = {Harewood,  F. J. and McHugh,  P. E.},
  year = {2007},
  month = may,
  pages = {1539–1553}
}

@article{Shanahan2017,
  title = {Viscoelastic braided stent: Finite element modelling and validation of crimping behaviour},
  volume = {121},
  ISSN = {0264-1275},
  url = {http://dx.doi.org/10.1016/j.matdes.2017.02.044},
  DOI = {10.1016/j.matdes.2017.02.044},
  journal = {Materials and Design},
  publisher = {Elsevier BV},
  author = {Shanahan,  Camelia and Tofail,  Syed A.M. and Tiernan,  Peter},
  year = {2017},
  month = may,
  pages = {143–153}
}

@article{Neffe2015,
  title = {One Step Creation of Multifunctional 3D Architectured Hydrogels Inducing Bone Regeneration},
  volume = {27},
  ISSN = {1521-4095},
  url = {http://dx.doi.org/10.1002/adma.201404787},
  DOI = {10.1002/adma.201404787},
  number = {10},
  journal = {Advanced Materials},
  publisher = {Wiley},
  author = {Neffe,  Axel T. and Pierce,  Benjamin F. and Tronci,  Giuseppe and Ma,  Nan and Pittermann,  Erik and Gebauer,  Tim and Frank,  Oliver and Schossig,  Michael and Xu,  Xun and Willie,  Bettina M. and Forner,  Michèle and Ellinghaus,  Agnes and Lienau,  Jasmin and Duda,  Georg N. and Lendlein,  Andreas},
  year = {2015},
  month = jan,
  pages = {1738–1744}
}

@article{Simon2003,
  title = {Early failure of the tissue engineered porcine heart valve SYNERGRAFT™ in pediatric patients},
  volume = {23},
  ISSN = {1010-7940},
  url = {http://dx.doi.org/10.1016/S1010-7940(03)00094-0},
  DOI = {10.1016/s1010-7940(03)00094-0},
  number = {6},
  journal = {European Journal of Cardio-Thoracic Surgery},
  publisher = {Oxford University Press (OUP)},
  author = {Simon,  P},
  year = {2003},
  month = jun,
  pages = {1002–1006}
}

@article{Jiao2012,
  title = {Measurements of the Effects of Decellularization on Viscoelastic Properties of Tissues in Ovine,  Baboon,  and Human Heart Valves},
  volume = {18},
  ISSN = {1937-335X},
  url = {http://dx.doi.org/10.1089/ten.tea.2010.0677},
  DOI = {10.1089/ten.tea.2010.0677},
  number = {3–4},
  journal = {Tissue Engineering Part A},
  publisher = {Mary Ann Liebert Inc},
  author = {Jiao,  Tong and Clifton,  Rodney J. and Converse,  Gabriel L. and Hopkins,  Richard A.},
  year = {2012},
  month = feb,
  pages = {423–431}
}

@article{Schapery1984,
  title = {Correspondence principles and a generalizedJ integral for large deformation and fracture analysis of viscoelastic media},
  volume = {25},
  ISSN = {1573-2673},
  url = {http://dx.doi.org/10.1007/BF01140837},
  DOI = {10.1007/bf01140837},
  number = {3},
  journal = {International Journal of Fracture},
  publisher = {Springer Science and Business Media LLC},
  author = {Schapery,  R. A.},
  year = {1984},
  month = jul,
  pages = {195–223}
}

@article{Rice1968,
  title = {A Path Independent Integral and the Approximate Analysis of Strain Concentration by Notches and Cracks},
  volume = {35},
  ISSN = {1528-9036},
  url = {http://dx.doi.org/10.1115/1.3601206},
  DOI = {10.1115/1.3601206},
  number = {2},
  journal = {Journal of Applied Mechanics},
  publisher = {ASME International},
  author = {Rice,  J. R.},
  year = {1968},
  month = jun,
  pages = {379–386}
}

@article{Graham1973,
  title = {The correspondence principle of linear viscoelasticity for problems that involve time-dependent regions},
  volume = {11},
  ISSN = {0020-7225},
  url = {http://dx.doi.org/10.1016/0020-7225(73)90074-8},
  DOI = {10.1016/0020-7225(73)90074-8},
  number = {1},
  journal = {International Journal of Engineering Science},
  publisher = {Elsevier BV},
  author = {Graham,  G.A.C. and Sabin,  G.C.W.},
  year = {1973},
  month = jan,
  pages = {123–140}
}

@article{Wang2012,
  title = {Delayed fracture in gels},
  volume = {8},
  ISSN = {1744-6848},
  url = {http://dx.doi.org/10.1039/c2sm25553g},
  DOI = {10.1039/c2sm25553g},
  number = {31},
  journal = {Soft Matter},
  publisher = {Royal Society of Chemistry (RSC)},
  author = {Wang,  Xiao and Hong,  Wei},
  year = {2012},
  pages = {8171}
}

@article{Guarino2002,
  title = {Failure time and critical behaviour of fracture precursors in heterogeneous materials},
  volume = {26},
  ISSN = {1434-6028},
  url = {http://dx.doi.org/10.1140/epjb/e20020075},
  DOI = {10.1140/epjb/e20020075},
  number = {2},
  journal = {The European Physical Journal B},
  publisher = {Springer Science and Business Media LLC},
  author = {Guarino,  A. and Ciliberto,  S. and Garcimartın,  A. and Zei,  M. and Scorretti,  R.},
  year = {2002},
  month = mar,
  pages = {141–151}
}

@article{Kun2003,
  title = {Creep rupture has two universality classes},
  volume = {63},
  ISSN = {1286-4854},
  url = {http://dx.doi.org/10.1209/epl/i2003-00469-9},
  DOI = {10.1209/epl/i2003-00469-9},
  number = {3},
  journal = {Europhysics Letters (EPL)},
  publisher = {IOP Publishing},
  author = {Kun,  F and Moreno,  Y and Hidalgo,  R. C and Herrmann,  H. J},
  year = {2003},
  month = aug,
  pages = {347–353}
}

@article{Ju2023,
  title = {Real-Time Early Detection of Crack Propagation Precursors in Delayed Fracture of Soft Elastomers},
  volume = {13},
  ISSN = {2160-3308},
  url = {http://dx.doi.org/10.1103/PhysRevX.13.021030},
  DOI = {10.1103/physrevx.13.021030},
  number = {2},
  journal = {Physical Review X},
  publisher = {American Physical Society (APS)},
  author = {Ju,  Jianzhu and Sanoja,  Gabriel E. and Nagazi,  Med Yassine and Cipelletti,  Luca and Liu,  Zezhou and Hui,  Chung Yuen and Ciccotti,  Matteo and Narita,  Tetsuharu and Creton,  Costantino},
  year = {2023},
  month = may 
}

@article{vanderKooij2018,
  title = {Laser Speckle Strain Imaging reveals the origin of delayed fracture in a soft solid},
  volume = {4},
  ISSN = {2375-2548},
  url = {http://dx.doi.org/10.1126/sciadv.aar1926},
  DOI = {10.1126/sciadv.aar1926},
  number = {5},
  journal = {Science Advances},
  publisher = {American Association for the Advancement of Science (AAAS)},
  author = {van der Kooij,  Hanne M. and Dussi,  Simone and van de Kerkhof,  Gea T. and Frijns,  Raoul A. M. and van der Gucht,  Jasper and Sprakel,  Joris},
  year = {2018},
  month = may 
}

@article{Grzelka2017,
  title = {Capillary fracture of ultrasoft gels: variability and delayed nucleation},
  volume = {13},
  ISSN = {1744-6848},
  url = {http://dx.doi.org/10.1039/C7SM00257B},
  DOI = {10.1039/c7sm00257b},
  number = {16},
  journal = {Soft Matter},
  publisher = {Royal Society of Chemistry (RSC)},
  author = {Grzelka,  Marion and Bostwick,  Joshua B. and Daniels,  Karen E.},
  year = {2017},
  pages = {2962–2966}
}

@article{Sprakel2011,
  title = {Stress Enhancement in the Delayed Yielding of Colloidal Gels},
  volume = {106},
  ISSN = {1079-7114},
  url = {http://dx.doi.org/10.1103/PhysRevLett.106.248303},
  DOI = {10.1103/physrevlett.106.248303},
  number = {24},
  journal = {Physical Review Letters},
  publisher = {American Physical Society (APS)},
  author = {Sprakel,  Joris and Lindstr\"{o}m,  Stefan B. and Kodger,  Thomas E. and Weitz,  David A.},
  year = {2011},
  month = jun 
}

@article{Frassine1996,
  title = {Experimental analysis of viscoelastic criteria for crack initiation and growth in polymers},
  volume = {81},
  ISSN = {1573-2673},
  url = {http://dx.doi.org/10.1007/BF00020755},
  DOI = {10.1007/bf00020755},
  number = {1},
  journal = {International Journal of Fracture},
  publisher = {Springer Science and Business Media LLC},
  author = {Frassine,  R. and Rink,  M. and Leggio,  A. and Pavan,  A.},
  year = {1996},
  pages = {55–75}
}

@article{Kaminsky2014,
  title = {Mechanics of the Delayed Fracture of Viscoelastic Bodies with Cracks: Theory and Experiment (Review)},
  volume = {50},
  ISSN = {1573-8582},
  url = {http://dx.doi.org/10.1007/s10778-014-0652-8},
  DOI = {10.1007/s10778-014-0652-8},
  number = {5},
  journal = {International Applied Mechanics},
  publisher = {Springer Science and Business Media LLC},
  author = {Kaminsky,  A. A.},
  year = {2014},
  month = sep,
  pages = {485–548}
}

@article{Wang2023,
  title = {Delayed fracture caused by time-dependent damage in PDMS},
  volume = {181},
  ISSN = {0022-5096},
  url = {http://dx.doi.org/10.1016/j.jmps.2023.105459},
  DOI = {10.1016/j.jmps.2023.105459},
  journal = {Journal of the Mechanics and Physics of Solids},
  publisher = {Elsevier BV},
  author = {Wang,  Jikun and Zhu,  Bangguo and Hui,  Chung-Yuen and Zehnder,  Alan T.},
  year = {2023},
  month = dec,
  pages = {105459}
}

@article{Ciavarella2021.2,
  title = {Crack propagation at the interface between viscoelastic and elastic materials},
  volume = {257},
  ISSN = {0013-7944},
  url = {http://dx.doi.org/10.1016/j.engfracmech.2021.108009},
  DOI = {10.1016/j.engfracmech.2021.108009},
  journal = {Engineering Fracture Mechanics},
  publisher = {Elsevier BV},
  author = {Ciavarella,  M. and Papangelo,  A. and McMeeking,  R.},
  year = {2021},
  month = nov,
  pages = {108009}
}

@article{Hui2022,
  title = {Steady state crack growth in viscoelastic solids: A comparative study},
  volume = {159},
  ISSN = {0022-5096},
  url = {http://dx.doi.org/10.1016/j.jmps.2021.104748},
  DOI = {10.1016/j.jmps.2021.104748},
  journal = {Journal of the Mechanics and Physics of Solids},
  publisher = {Elsevier BV},
  author = {Hui,  Chung-Yuen and Zhu,  Bangguo and Long,  Rong},
  year = {2022},
  month = feb,
  pages = {104748}
}

@article{DasGupta1977,
  title = {Delayed Yielding of a Plane Stress Viscoelastic Dugdale Model},
  volume = {5},
  ISSN = {0090-3973},
  url = {http://dx.doi.org/10.1520/JTE10556J},
  DOI = {10.1520/jte10556j},
  number = {6},
  journal = {Journal of Testing and Evaluation},
  publisher = {ASTM International},
  author = {DasGupta,  A and Brinson,  HF},
  year = {1977},
  month = nov,
  pages = {437–445}
}

@article{Ciavarella2022,
  title = {Viscoelastic short cracks propagation},
  volume = {264},
  ISSN = {0013-7944},
  url = {http://dx.doi.org/10.1016/j.engfracmech.2022.108276},
  DOI = {10.1016/j.engfracmech.2022.108276},
  journal = {Engineering Fracture Mechanics},
  publisher = {Elsevier BV},
  author = {Ciavarella,  M.},
  year = {2022},
  month = apr,
  pages = {108276}
}

@article{Irwin1957,
  title = {Analysis of Stresses and Strains Near the End of a Crack Traversing a Plate},
  volume = {24},
  ISSN = {1528-9036},
  url = {http://dx.doi.org/10.1115/1.4011547},
  DOI = {10.1115/1.4011547},
  number = {3},
  journal = {Journal of Applied Mechanics},
  publisher = {ASME International},
  author = {Irwin,  G. R.},
  year = {1957},
  month = sep,
  pages = {361–364}
}

@article{Tabuteau2009,
  title = {Microscopic Mechanisms of the Brittleness of Viscoelastic Fluids},
  volume = {102},
  ISSN = {1079-7114},
  url = {http://dx.doi.org/10.1103/PhysRevLett.102.155501},
  DOI = {10.1103/physrevlett.102.155501},
  number = {15},
  journal = {Physical Review Letters},
  publisher = {American Physical Society (APS)},
  author = {Tabuteau,  H. and Mora,  S. and Porte,  G. and Abkarian,  M. and Ligoure,  C.},
  year = {2009},
  month = apr 
}

@article{Evans1997,
  title = {Dynamic strength of molecular adhesion bonds},
  volume = {72},
  ISSN = {0006-3495},
  url = {http://dx.doi.org/10.1016/S0006-3495(97)78802-7},
  DOI = {10.1016/s0006-3495(97)78802-7},
  number = {4},
  journal = {Biophysical Journal},
  publisher = {Elsevier BV},
  author = {Evans,  E. and Ritchie,  K.},
  year = {1997},
  month = apr,
  pages = {1541–1555}
}

@phdthesis{Mueller,
  doi = {10.7907/MGSD-R362},
  url = {https://resolver.caltech.edu/CaltechETD:etd-11152005-142505},
  author = {Mueller,  Hans-Karl Christian Alfred},
  language = {en},
  title = {Stable crack propagation in a viscoelastic strip},
  school = {California Institute of Technology},
  year = {1968},
  copyright = {No commercial reproduction,  distribution,  display or performance rights in this work are provided.}
}

@article{Carbone2005,
  title = {Crack motion in viscoelastic solids: The role of the flash temperature},
  volume = {17},
  ISSN = {1292-895X},
  url = {http://dx.doi.org/10.1140/epje/i2005-10013-y},
  DOI = {10.1140/epje/i2005-10013-y},
  number = {3},
  journal = {The European Physical Journal E},
  publisher = {Springer Science and Business Media LLC},
  author = {Carbone,  G. and Persson,  B. N. J.},
  year = {2005},
  month = jul,
  pages = {261–281}
}

@article{Onsager1931,
  title = {Reciprocal Relations in Irreversible Processes. I.},
  volume = {37},
  ISSN = {0031-899X},
  url = {http://dx.doi.org/10.1103/PhysRev.37.405},
  DOI = {10.1103/physrev.37.405},
  number = {4},
  journal = {Physical Review},
  publisher = {American Physical Society (APS)},
  author = {Onsager,  Lars},
  year = {1931},
  month = feb,
  pages = {405–426}
}

@book{Callen1985,
  author    = {Herbert B. Callen},
  title     = {Thermodynamics and an Introduction to Thermostatistics},
  edition   = {2},
  year      = {1985},
  publisher = {John Wiley \& Sons},
  address   = {New York}
}

@inbook{Maugis2000,
  title = {Rupture and Adherence of Elastic Solids},
  ISBN = {9783662041253},
  ISSN = {0171-1873},
  url = {http://dx.doi.org/10.1007/978-3-662-04125-3_3},
  DOI = {10.1007/978-3-662-04125-3_3},
  booktitle = {Contact,  Adhesion and Rupture of Elastic Solids},
  publisher = {Springer Berlin Heidelberg},
  author = {Maugis,  Daniel},
  year = {2000},
  pages = {133–202}
}

@article{Onsager1931.2,
  title = {Reciprocal Relations in Irreversible Processes. II.},
  volume = {38},
  ISSN = {0031-899X},
  url = {http://dx.doi.org/10.1103/PhysRev.38.2265},
  DOI = {10.1103/physrev.38.2265},
  number = {12},
  journal = {Physical Review},
  publisher = {American Physical Society (APS)},
  author = {Onsager,  Lars},
  year = {1931},
  month = dec,
  pages = {2265–2279}
}

@article{Mori1965,
  title = {Transport,  Collective Motion,  and Brownian Motion},
  volume = {33},
  ISSN = {0033-068X},
  url = {http://dx.doi.org/10.1143/PTP.33.423},
  DOI = {10.1143/ptp.33.423},
  number = {3},
  journal = {Progress of Theoretical Physics},
  publisher = {Oxford University Press (OUP)},
  author = {Mori,  Hazime},
  year = {1965},
  month = mar,
  pages = {423–455}
}

@article{Kubo1966,
  title = {The fluctuation-dissipation theorem},
  volume = {29},
  ISSN = {0034-4885},
  url = {http://dx.doi.org/10.1088/0034-4885/29/1/306},
  DOI = {10.1088/0034-4885/29/1/306},
  number = {1},
  journal = {Reports on Progress in Physics},
  publisher = {IOP Publishing},
  author = {Kubo,  R},
  year = {1966},
  month = jan,
  pages = {255–284}
}

@article{Carbone2005.2,
  title = {Hot Cracks in Rubber: Origin of the Giant Toughness of Rubberlike Materials},
  volume = {95},
  ISSN = {1079-7114},
  url = {http://dx.doi.org/10.1103/PhysRevLett.95.114301},
  DOI = {10.1103/physrevlett.95.114301},
  number = {11},
  journal = {Physical Review Letters},
  publisher = {American Physical Society (APS)},
  author = {Carbone,  G. and Persson,  B. N. J.},
  year = {2005},
  month = sep 
}

@article{Zhang2023,
  title = {Pressure and polymer selections for solid-state batteries investigated with high-throughput simulations},
  volume = {4},
  ISSN = {2666-3864},
  url = {http://dx.doi.org/10.1016/j.xcrp.2023.101328},
  DOI = {10.1016/j.xcrp.2023.101328},
  number = {3},
  journal = {Cell Reports Physical Science},
  publisher = {Elsevier BV},
  author = {Zhang,  Xin and Luo,  Changqi and Menga,  Nicola and Zhang,  Hao and Li,  Yanxin and Zhu,  Shun-Peng},
  year = {2023},
  month = mar,
  pages = {101328}
}

@article{Cheng2024,
  title = {Interfacial performance evolution of ceramics-in-polymer composite electrolyte in solid-state lithium metal batteries},
  volume = {204},
  ISSN = {0020-7225},
  url = {http://dx.doi.org/10.1016/j.ijengsci.2024.104137},
  DOI = {10.1016/j.ijengsci.2024.104137},
  journal = {International Journal of Engineering Science},
  publisher = {Elsevier BV},
  author = {Cheng,  Ao and Sun,  Linlin and Menga,  Nicola and Yang,  Wanyou and Zhang,  Xin},
  year = {2024},
  month = nov,
  pages = {104137}
}

@article{Ding2021,
  title = {A review on the failure and regulation of solid electrolyte interphase in lithium batteries},
  volume = {59},
  ISSN = {2095-4956},
  url = {http://dx.doi.org/10.1016/j.jechem.2020.11.016},
  DOI = {10.1016/j.jechem.2020.11.016},
  journal = {Journal of Energy Chemistry},
  publisher = {Elsevier BV},
  author = {Ding,  Jun-Fan and Xu,  Rui and Yan,  Chong and Li,  Bo-Quan and Yuan,  Hong and Huang,  Jia-Qi},
  year = {2021},
  month = aug,
  pages = {306–319}
}

@article{Nilsson1973,
  author  = {Fred Nilsson},
  title   = {A path-independent integral for transient crack problems},
  journal = {International Journal of Solids and Structures},
  year    = {1973},
  volume  = {9},
  number  = {9},
  pages   = {1107--1115},
  doi     = {10.1016/0020-7683(73)90009-8}
}

@book{tada2000stress,
  title     = {The Stress Analysis of Cracks Handbook},
  edition   = {3},
  author    = {Tada, Hiroshi and Paris, Paul C. and Irwin, George R.},
  year      = {2000},
  publisher = {ASME Press},
  isbn      = {0791801535},
  doi       = {10.1115/1.801535}
}

@article{Joyce2003PTFE,
  author    = {James A. Joyce},
  title     = {Fracture Toughness Evaluation of Polytetrafluoroethylene},
  journal   = {Polymer Engineering and Science},
  year      = {2003},
  volume    = {43},
  number    = {10},
  pages     = {1702--1714},
  doi       = {10.1002/pen.10144}
}

@article{JoyceJoyce2004PTFE,
  author    = {Peter J. Joyce and James A. Joyce},
  title     = {Evaluation of the Fracture Toughness Properties of Polytetrafluoroethylene},
  journal   = {International Journal of Fracture},
  year      = {2004},
  volume    = {127},
  pages     = {361--385},
  doi       = {10.1023/B:FRAC.0000037674.46965.FB}
}

\end{document}